\documentclass[10pt, twocolumn, comsoc]{IEEEtran}

\usepackage{graphicx,epsfig}
\usepackage[noadjust]{cite}
\usepackage{mcite}
\usepackage{amsfonts,helvet}
\usepackage{fancyhdr}
\usepackage{threeparttable}
\usepackage{epsf,epsfig}
\usepackage{amsthm}
\usepackage{amsmath}
\usepackage{siunitx}
\usepackage{amssymb}
\usepackage{stfloats}

\usepackage{dsfont}
\usepackage{subfigure}
\usepackage{color}
\usepackage{enumerate}
\usepackage{gensymb}
\usepackage{cancel}
\usepackage{lipsum}
\usepackage{mathtools}
\usepackage{cuted}
\usepackage{bbm}
\usepackage[linesnumbered,ruled]{algorithm2e}

\newtheorem{remark}{Remark}

\usepackage[colorlinks=true, linkcolor=blue]{hyperref}

\usepackage{eucal}

\begin{document}

\title{Spatial MIMO Channel Rendering: \\ A 3D Gaussian Splatting Approach}

\author{Juntaek~Han, Geonha~Kim, Chaozheng~Wen, Ahmed~Alkhateeb, Jun~Zhang, and Jeonghun Park

\thanks{This work was supported in part by the Institute of Information \& communications Technology Planning \& Evaluation (IITP) grant funded by the Korea government (MSIT) (No. RS-2024-00397216, Development of the Upper-mid Band Extreme massive MIMO (E-MIMO), No. RS-2024-00395824, Development of Cloud virtualized RAN (vRAN) system supporting upper-midband, and No. RS-2026-25523842), and in part by the 6GARROW project which has received funding from the Smart Networks and Services Joint Undertaking (SNS JU) under the European Union's Horizon Europe research and innovation programme (Grant Agreement No. 101192194) and from IITP (No. RS-2024-00435652).

J. Han, G. Kim, and J. Park are with School of Electrical and Electronic Engineering, Yonsei University, Seoul, South Korea (e-mail:{\texttt{jthan1218@yonsei.ac.kr, drawnemo@yonsei.ac.kr, jhpark@yonsei.ac.kr}}).

A. Alkhateeb is with the School of Electrical, Computer and Energy Engineering, Arizona State University, Tempe, USA (e-mail: {\texttt{alkhateeb@asu.edu}}).

C. Wen and J. Zhang are with the Department of Electronic and Computer Engineering, The Hong Kong University of Science and Technology, Hong Kong (e-mail: {\texttt{cwenae@connect.ust.hk, eejzhang@ust.hk}}).

The implementation and the dataset are available at \url{https://github.com/jthan1218/MIMO-GS}.
}}

\maketitle \setcounter{page}{1}

\begin{abstract}
    The achievable gains of multiple-input multiple-output (MIMO) systems depend on channel state information (CSI), whose acquisition consumes pilot and feedback resources for every user in every coherence interval. Meanwhile, the propagation environment that shapes the channel persists. Exploiting such information will help to greatly reduce the overhead of CSI acquisition. This paper investigates the \textit{spatial MIMO channel rendering} problem, which learns a latent beamspace channel scene from sparse observations and renders the long-term transmit and receive beam pair power map at unobserved user locations. Motivated by 3D Gaussian splatting in visual scene rendering, the proposed MIMO Gaussian splatting (MIMO-GS) represents the scene with pairs of receive and transmit 3D Gaussians, projects them from the user and base station onto the corresponding beam domains, and accumulates the beam associations through beam pair splatting. On a simulated urban scenario and measured indoor channels, MIMO-GS renders $5.5$ times faster than ray tracing without a calibrated replica. Applied to multiuser beam management, the rendered maps allow users and transmit beams to be scheduled without sounding while accounting for the inter-user interference, and the net sum rate reaches $98\%$ of the genie-aided benchmark, $19\%$ above a position-based predictor.
\end{abstract}


\section{Introduction}




Multiple-input multiple-output (MIMO) technology has become a cornerstone of modern wireless systems by offering spatial multiplexing, array, and diversity gains. The achievable gains in practice, however, heavily depend on the available channel state information (CSI) at the transmitter. Without accurate CSI, precoding is misdirected, inter-user interference remains uncontrolled, and even the multiplexing gain of the multiuser downlink is lost \cite{park16feedback, jindal2006finite}. CSI is thus the key ingredient from which the MIMO gains are extracted, but its acquisition can be very costly. It consumes both pilot and feedback resources \cite{hassibi2003howmuch, love2008overview}, which grow with the number of antennas. Moreover, since CSI expires after the coherence time, the cost must be paid again for every user in every coherence interval. Efficient CSI acquisition is therefore central to realizing the MIMO gains in practice.

A basic form of CSI acquisition follows an estimate-then-feedback mechanism. For instance, in a frequency division duplex (FDD) downlink where uplink-downlink reciprocity does not hold, the base station (BS) transmits downlink pilots, and the user equipment (UE) estimates the downlink channel from these pilots; then the UE reports the estimate back to the BS by quantizing it with a predefined codebook \cite{love2008overview, park16feedback}. This so-called \emph{estimation} mechanism has served as the standard approach to MIMO channel acquisition throughout the literature. The current cellular standards are also built on it, including the CSI framework for 5G New Radio (NR) \cite{3gpp38214}. A fundamental limit of this mechanism is its poor scalability. The pilot overhead grows with the spatial dimension of the channel, while the feedback overhead grows accordingly, multiplied by the number of users and frequency resources.



To overcome the inefficiency of pilot-based channel estimation, several studies have explored the idea of {\emph{prediction}}. Its key principle is that channels already acquired serve as side information for an unobserved one. The channel at an unobserved index can then be recovered, at least in part, without additional pilots.
A representative example is time-domain prediction. Because the channel varies smoothly over time rather than independently from one instant to the next, a future channel can be predicted from past estimates. Classical methods model this temporal evolution with autoregressive or Kalman filtering \cite{baddour2005autoregressive}, while more recent ones learn it with deep sequence models \cite{wu2021channel}. 

More recently, prediction across the frequency axis has also been actively studied, where the channel acquired in one band is used to infer the channel in another. For instance, in an FDD system, the uplink and downlink occupy separate frequency bands, so their instantaneous channels differ; yet both arise from the same physical propagation and share the same paths, angles, and delays. This frequency-invariant structure, known as partial reciprocity, allows the downlink channel to be inferred from uplink measurements. This idea has been pursued by estimating the reciprocal angles and delays on the uplink and extrapolating them to reconstruct the downlink channel \cite{Rottenberg:2020}, as well as by learning the uplink-to-downlink mapping directly with deep neural networks \cite{alhed19asilomar}. Pushing this principle further, a recent line of work eliminates the dedicated downlink pilots and the associated CSI feedback altogether, obtaining from uplink pilots both the downlink channel and its error covariance and designing the precoder directly from them \cite{kim2025splitting, han2026reducing, kim2026integrated}.

Compared to the time and frequency domains, prediction across the \emph{space} domain remains largely unexplored. Spatial channel prediction refers to inferring the channel at an unobserved location from channels acquired at other locations. Its feasibility stems from the shared physical environment that determines the channel propagation across different locations. Since the channel is the response of a given environment, the channels at different locations are related, as nearby points see a common set of scatterers and are reached by overlapping propagation paths. Spatial prediction enjoys two advantages. First, the environment is persistent, whereas the instantaneous channel expires after the coherence time. The structure induced by the environment can therefore be reused across channel realizations and users as long as the propagation scene and the operating configuration remain unchanged. Second, spatial prediction can reveal the channel of a user that has never been estimated before \cite{va2019}.



The spatial domain, however, is far less amenable to such prediction. 
Temporal and spectral predictions rest on simple, explicit models. A channel correlates with its recent past, and its angles and delays carry over to a nearby band, so the channel at an unobserved index follows from a compact relation. Space admits no such shortcut. To know the channel at an unobserved location, one must understand how it is shaped by the surrounding geometry including the scatterers, reflectors, and blockages, a dependence that no simple model captures. 

This raises the question of what a reusable description of the propagation environment should contain. Instantaneous CSI characterizes a particular channel realization and expires with it, whereas the distribution of channel power over propagation directions persists. Since each propagation component couples a departure direction at the transmitter with an arrival direction at the receiver, this directional structure has to be kept jointly over the two sides. We therefore take the long-term joint transmit and receive beam pair power map as the prediction target. We refer to this approach as channel \emph{rendering}, in analogy to computer graphics, where a scene representation is learned once and an image is rendered from an arbitrary viewpoint \cite{renderingsurvey}. In our setting, the scene is the MIMO beamspace structure induced by the environment, and it is learned from a finite set of channel observations. The viewpoint is the UE location, and the rendered output is the beam pair power map at locations where no measurement has been taken.

\begin{table*}[t]
\centering
\caption{Comparison of environment-based channel rendering approaches}
\label{table_related_works}
\setlength{\tabcolsep}{4pt}
\begin{threeparttable}
\begin{tabular*}{\textwidth}{@{\extracolsep{\fill}} l l l l l @{}}
\hline
Approach & Scene representation & Rendering domain & Rendered output & Antennas \\
\hline
Ray tracing \cite{yun2015raytracing, hoydis2023sionna} & Physical 3D replica & Propagation scene & Propagation paths and channels & Arbitrary \\
NeRF$^2$ \cite{nerf2} & NeRF & Rx angular domain & Receive spatial spectrum & Single antenna Tx \\
NeWRF \cite{newrf} & NeRF & Rx angular domain & SISO channel coefficient & Single antenna Tx \\
RF-3DGS \cite{rf3dgs} & 3D-GS & Rx angular domain & Receive spatial spectrum\tnote{a} & Single antenna Tx \\
WRF-GS+ \cite{wrfgs} & 3D-GS & Rx angular domain & Receive spatial spectrum and CSI & Single antenna Tx \\
BiWGS \cite{biwgs} & 3D-GS\tnote{b} & Rx angular domain & SIMO channel vector & Single antenna Tx \\
GSBF \cite{gsbf} & 3D-GS & Tx angular domain & Beamforming vector & Single antenna Rx \\
\textbf{MIMO-GS (proposed)} & \textbf{Dual-sided 3D-GS} & \textbf{MIMO beamspace} & \textbf{Long-term MIMO beamspace map} & \textbf{Arrays at both ends} \\
\hline
\end{tabular*}
\begin{tablenotes}\footnotesize
\item[a] The angle of departure and the delay are attached to the spectrum as rendered attributes of each path, not as output dimensions.
\item[b] The channel vector is rendered per Tx--Rx position pair, with both endpoint positions taken as inputs.
\end{tablenotes}
\end{threeparttable}
\end{table*}

\subsection{Related Works}

The objective of spatial channel rendering is to obtain site-specific channel information at locations where no measurement has been taken. A conventional approach to this objective is ray tracing, which models the channel at an arbitrary location by simulating electromagnetic propagation over a 3D model of the environment \cite{yun2015raytracing, hoydis2023sionna} and serves as the core engine of the recently envisioned wireless digital twins \cite{alkhateeb2023digitaltwin}. Ideally, it offers the same capability that we pursue, namely site-specific channel knowledge at locations where no measurement has been taken.
Its accuracy, however, is bounded by the fidelity of the underlying digital replica. The simulation requires the geometry, material properties, and electromagnetic parameters of every surface in the scene, none of which is exactly known, and any mismatch between the replica and the real environment propagates directly into the predicted channel \cite{jiang2023dtbeam}. High fidelity simulation is also computationally expensive, since the cost scales with the complexity of the site and the number of propagation interactions.

To reduce the dependence on a fully calibrated physical replica, recent radio frequency (RF) studies have learned continuous representations of the radio scene directly from channel observations. This line of work builds on neural radiance fields (NeRF)  \cite{nerf}, originally developed in computer vision to represent a three-dimensional scene as an implicit neural function and synthesize observations by evaluating the field along viewing rays. NeRF$^2$ \cite{nerf2} and NeWRF \cite{newrf} extend this framework to radio propagation by learning continuous neural fields from channel observations, enabling RF quantities or channel responses to be synthesized at unobserved locations. Their representation remains implicit, however, and rendering typically requires repeated neural network evaluations over many spatial samples or rays, which can impose substantial training and inference cost.

An explicit alternative to these implicit neural fields is 3D Gaussian splatting (3D-GS) \cite{kerbl:acm:23}, which was originally developed for visual scene rendering. It represents a scene with explicit anisotropic Gaussian primitives that are projected and accumulated on an image plane, which allows efficient scene optimization and rendering. Recent studies have adapted this representation to radio propagation by redefining the Gaussian attributes, the projection geometry, and the aggregation rule. RF-3DGS \cite{rf3dgs} renders radio spatial spectra and the gain, delay, angle of arrival, and angle of departure of each path. WRF-GS \cite{wrfgs_infocom} and WRF-GS+ \cite{wrfgs} reconstruct wireless radiation fields and spatial spectra and separately evaluate CSI prediction from the uplink to the downlink, while GSRF \cite{gsrf} employs Gaussians with complex-valued attributes to synthesize RF data with phase. More recent Gaussian field formulations broaden the scope further. In nGRF \cite{ngrf}, complex-valued Gaussian contributions are aggregated directly in 3D space for channel estimation, whereas BiWGS \cite{biwgs} constructs channel maps over six dimensions spanning the transmitter and receiver positions. GSBF \cite{gsbf} renders a complex-valued spectrum over the transmit angles of the BS and combines it with the array response vectors to form the beamforming vector toward a single-antenna user.

As summarized in Table~\ref{table_related_works}, a common limitation of the existing works is that they build on a link with a single antenna at one end, so the rendered quantity is a spectrum or channel observed from one side alone. Such one-sided formulations cannot be applied directly to the MIMO beamspace targeted here, where each propagation component couples a transmit beam with a receive beam. Rendering the joint beam pair power map therefore calls for a projection and aggregation mechanism that spans both beam domains.

\subsection{Contributions}

To address this gap, we propose MIMO Gaussian splatting (MIMO-GS), a 3D-GS framework for rendering the MIMO beamspace channel scene. Instead of building a replica of the environment and simulating propagation over it, MIMO-GS learns an explicit latent scene directly from channel observations and reduces rendering to a lightweight inference step. The rendered long-term map then serves as the statistical CSI for multiuser transmission in the beam domain, so that CSI reference signal (CSI-RS) resources are configured only on the beams scheduled from the map.

The main contributions of this work are summarized as follows:
\begin{itemize}
    \item \textbf{Rendering MIMO beamspace channel scene}: we formulate spatial MIMO channel rendering as learning a continuous long-term beamspace scene from channel observations, with the joint transmit and receive beam pair power map as the rendered quantity, which preserves the coupled angular structure of the MIMO channel.
    \item \textbf{Dual-sided projection for MIMO beamspace rendering}: we design the rendering mechanism of MIMO-GS, in which each primitive pairs a receive Gaussian and a transmit Gaussian, projected from the UE and the BS onto the corresponding beam domains. A gain conditioned on the UE location scales the contribution of each primitive, and a differentiable beam pair splatting accumulates the contributions by direct nonnegative aggregation. The rendering is validated on a ray-traced urban scenario and over-the-air measured indoor channels.
    \item \textbf{Multiuser beam management based on the rendered map}: we develop a beam management scheme in which the rendered maps replace the statistical CSI that multiuser transmission would otherwise obtain by sounding. Users and transmit beams are allocated from the maps under the inter-user interference they create, and CSI-RS resources are configured only on the scheduled beams. The scheme is evaluated with the net sum rate, which discounts the spectral efficiency by this CSI-RS overhead.
\end{itemize}

The rest of this paper is organized as follows. Section II formulates the spatial channel rendering problem, and Section III reviews 3D Gaussian splatting. Section IV presents the proposed MIMO-GS rendering framework, and Section V the beam management scheme on the rendered map. Section VI provides numerical results, and Section VII concludes this paper.

\emph{Notation}: Boldface lowercase and uppercase letters denote vectors and matrices. $\mathbb{R}^{m\times n}$ and $\mathbb{C}^{m\times n}$ denote the spaces of real and complex $m\times n$ matrices, and $\mathbb{R}_+$ the nonnegative reals. $(\cdot)^{\mathsf{T}}$, $(\cdot)^{\mathsf{H}}$, and $(\cdot)^{-1}$ denote the transpose, Hermitian transpose, and inverse, $\mathbf{I}_N$ the $N\times N$ identity matrix, $\|\cdot\|_{\mathrm{F}}$ the Frobenius norm, $\|\cdot\|_2$ the Euclidean norm, $|\mathcal{A}|$ the cardinality of a set $\mathcal{A}$, and $\mathbb{E}[\cdot]$ the expectation.

\section{Channel Model and Problem Formulation} \label{section_channel_model}

We consider a downlink MIMO channel in a three-dimensional wireless environment, where a fixed BS serves a UE whose location varies within the environment. The BS and UE are equipped with uniform planar arrays (UPAs). The BS UPA has \(N_{\mathrm{t},\mathrm{h}}\) horizontal and \(N_{\mathrm{t},\mathrm{v}}\) vertical elements, whereas the UE UPA has \(N_{\mathrm{r},\mathrm{h}}\) horizontal and \(N_{\mathrm{r},\mathrm{v}}\) vertical elements. Hence, the numbers of transmit and receive antennas are \(N_\mathrm{t}=N_{\mathrm{t},\mathrm{h}}N_{\mathrm{t},\mathrm{v}}\) and \(N_\mathrm{r}=N_{\mathrm{r},\mathrm{h}}N_{\mathrm{r},\mathrm{v}}\), respectively. For a UE located at $\mathbf{p}\in\mathbb{R}^3$, the propagation geometry induced by the BS and surrounding scatterers determines the effective multipath components. Accordingly, the set of signal paths, their propagation distances, complex path gains, and angular parameters are all functions of the UE location $\mathbf{p}$.
Hence, we model the downlink MIMO channel as a location-dependent matrix $\mathbf{H}(\mathbf{p})\in\mathbb{C}^{N_\mathrm{r}\times N_{\mathrm{t}}}$, which defines a MIMO channel field $\mathbf{p}\mapsto\mathbf{H}(\mathbf{p})$. 
Specifically, with an angular domain multipath representation, the channel is written as 
\begin{align}
    \mathbf{H}(\mathbf{p})=\sum_{\ell\in\mathcal{R}(\mathbf{p})}\beta_\ell(\mathbf{p})\mathbf{a} _\mathrm{r}\left(\boldsymbol{\Omega}^{\mathrm{r}}_\ell(\mathbf{p})\right)\mathbf{a}_\mathrm{t}^{\sf H}\left(\boldsymbol{\Omega}^{\mathrm{t}}_\ell(\mathbf{p})\right),\label{2_channel_model}
\end{align}
where $\mathcal{R}(\mathbf{p})$ denotes the set of effective propagation paths observed at location $\mathbf{p}$, $\beta_\ell(\mathbf{p})$ is the complex coefficient of the $\ell$-th path, $\mathbf{a}_\mathrm{r}(\cdot)$ and $\mathbf{a}_\mathrm{t}(\cdot)$ are the receive and transmit UPA response vectors, and $\boldsymbol{\Omega}^{\mathrm{r}}_\ell(\mathbf{p})$ and $\boldsymbol{\Omega}^{\mathrm{t}}_\ell(\mathbf{p})$ denote the corresponding receive and transmit angular parameters, each measured from its own array toward the direction of the path. Under this effective narrowband representation, $\beta_\ell(\mathbf{p})$ includes the path gain, propagation phase, and any frequency-dependent phase factor with the considered frequency resource.


Since each path in \eqref{2_channel_model} is associated with both transmit and receive angular parameters, we represent the channel in the beamspace domain as
\begin{align}
    \mathbf{H}_{\mathrm{b}}(\mathbf{p}) = \mathbf{A}_\mathrm{r}^{\sf H} \mathbf{H}(\mathbf{p}) \mathbf{A}_\mathrm{t} \in \mathbb{C}^{N_{\mathrm{r}}\times N_{\mathrm{t}}},
\end{align}
where $\mathbf{A}_\mathrm{t}\in\mathbb{C}^{N_\mathrm{t}\times N_\mathrm{t}}$ and $\mathbf{A}_\mathrm{r}\in\mathbb{C}^{N_\mathrm{r}\times N_\mathrm{r}}$ denote the transmit and receive unitary discrete Fourier transform (DFT) beam matrices associated with the BS and UE UPAs, respectively. The $(m,n)$-th entry of $\mathbf{H}_{\mathrm{b}}(\mathbf{p})$, i.e., $[\mathbf{H}_{\mathrm{b}}(\mathbf{p})]_{m,n}$, represents the effective channel response for the beam pair consisting of the \(m\)-th receive beam and the \(n\)-th transmit beam. Therefore, the beamspace representation organizes the multipath channel at any given location into a structured domain of beam pairs. 

From the complex beamspace channel $\mathbf{H}_{\mathrm{b}}(\mathbf{p})$, we define the expected beam pair power map
$\mathbf{X}(\mathbf{p})\in\mathbb{R}_{+}^{N_{\mathrm{r}}\times N_{\mathrm{t}}}$. For $m=1,\ldots,N_{\mathrm{r}}$ and $n=1,\ldots,N_{\mathrm{t}}$, its $(m,n)$-th entry is given by
\begin{align}
    [\mathbf{X}(\mathbf{p})]_{m,n} \triangleq \mathbb{E}\left[ \left| [\mathbf{H}_{\mathrm{b}}(\mathbf{p})]_{m,n} \right|^{2} \right],\label{2_power_map}
\end{align}
where the expectation is taken over channel observations at the same UE location while the macroscopic propagation scene remains unchanged. 
Here, $[\mathbf{X}(\mathbf{p})]_{m,n}$ denotes the long-term power gain of the beam pair formed by the $m$-th receive and $n$-th transmit DFT beams.
Throughout the paper, the UPA beam coordinates are represented in the same two-dimensional coordinate domain used by the UPA array response vectors $\mathbf{a}_{\mathrm{r}}(\cdot)$ and $\mathbf{a}_{\mathrm{t}}(\cdot)$. 

\begin{remark} \normalfont
    The beam pair power map in \eqref{2_power_map} is the joint form of the quantity that is measured and reported in 3GPP systems, namely the received power of each beam obtained by sweeping a codebook \cite{3gpp38214}. A rendered map thus provides this quantity at unobserved locations without a sweep and can reduce the channel acquisition overhead. The map is also the part of the channel that the macroscopic environment determines, which is what spatial rendering recovers. The phase of each path in \eqref{2_channel_model} rotates through a full cycle whenever the path length changes by one wavelength, so the complex channel decorrelates within a wavelength of displacement and differs across realizations even at a fixed location. The beam pair powers, in contrast, are set by the path angles and gains and vary on the scale of the propagation geometry. The map at an unobserved location can therefore be recovered from observations that lie meters apart, which is the spacing of channel observations in practice.
\end{remark}

We next formulate the channel rendering problem based on the beamspace channel model. Specifically, we adopt the beam pair power map defined in \eqref{2_power_map} as the rendering target. This map describes the channel power gains over transmit and receive DFT beam pairs and thus captures how the dominant beam pairs vary across UE locations. Let $\mathcal{P}\subset\mathbb{R}^3$ denote the UE location domain considered for rendering. For $\mathbf{p}\in\mathcal{P}$, the desired rendering output is $\mathbf{X}(\mathbf{p})\in\mathbb{R}_+^{N_{\mathrm{r}}\times N_{\mathrm{t}}}$.

Assume that channel observations are available at $N$ UE locations, which form the observation set $\mathcal{D}=\{(\mathbf{p}_i,\mathbf{X}_i)\}_{i=1}^N$ with $\mathbf{X}_i=\mathbf{X}(\mathbf{p}_i)$. Given $\mathcal{D}$, our goal is to learn a parametric renderer $f_\Theta$ that approximates the underlying beam pair power scene:
\begin{align}
    \widehat{\mathbf{X}}(\mathbf{p})=f_\Theta(\mathbf{p}), \qquad f_\Theta: \mathcal{P}\rightarrow \mathbb{R}_+^{N_{\mathrm{r}}\times N_{\mathrm{t}}}.
\end{align}
Since the BS position and the transmit and receive DFT beam bases are fixed for the considered scene, they are omitted from the argument of $f_\Theta$.

The renderer is trained by minimizing a beam pair power map reconstruction loss over the available observations:
\begin{align}
    \Theta^\star=\arg\min_{\Theta}\frac{1}{N}\sum_{i=1}^N\mathcal{L}\left(f_\Theta(\mathbf{p}_i), \mathbf{X}_i\right),
\end{align}
where $\mathcal{L}(\cdot, \cdot)$ measures the discrepancy between the rendered and observed beam pair power maps.
After training, the map is rendered as $\widehat{\mathbf{X}}(\mathbf{p}^\star)=f_{\Theta^\star}(\mathbf{p}^\star)$ for an arbitrary UE location $\mathbf{p}^\star\in\mathcal{P}$, which refers to any location within this learned scene and does not imply extrapolation to a different propagation environment.

Reproducing the beam pair power map at unobserved locations from the $N$ observations alone, however, is far from straightforward. The first challenge lies in how the maps at different locations are related. As shown in \eqref{2_channel_model}, the entries of $\mathbf{X}(\mathbf{p})$ vary gradually as the angular parameters drift while $\mathcal{R}(\mathbf{p})$ remains unchanged. Crossing a blockage or visibility boundary, however, changes $\mathcal{R}(\mathbf{p})$ itself and can shift the dominant beam pairs abruptly even between nearby locations. Accordingly, rather than assuming a relation among the observed maps over $\mathcal{P}$, the renderer should represent the underlying scene induced by the environment, from which the map at any location can be rendered.

The second challenge arises from the structure of the rendering target itself. Since each propagation component in \eqref{2_channel_model} couples a departure direction $\boldsymbol{\Omega}^{\mathrm{t}}_\ell(\mathbf{p})$ at the BS with an arrival direction $\boldsymbol{\Omega}^{\mathrm{r}}_\ell(\mathbf{p})$ at the UE, the renderer must preserve this coupling over the $N_{\mathrm{r}}\times N_{\mathrm{t}}$ beam pair grid rather than modeling the two beam domains independently.

These challenges call for a scene representation that is continuous over $\mathcal{P}$, differentiable, capable of expressing both smooth trends and sharp spatial transitions, and structured over the joint beam pair domain. To this end, we build on 3D-GS, an explicit and differentiable scene representation reviewed in Section~\ref{3dgs}, and develop the proposed MIMO-GS rendering framework in Section~\ref{section_mimogs_framework}.


\section{Preliminaries: 3D Gaussian Splatting}\label{3dgs}

\begin{figure*}[t]
    \centering
    \includegraphics[width=\textwidth]{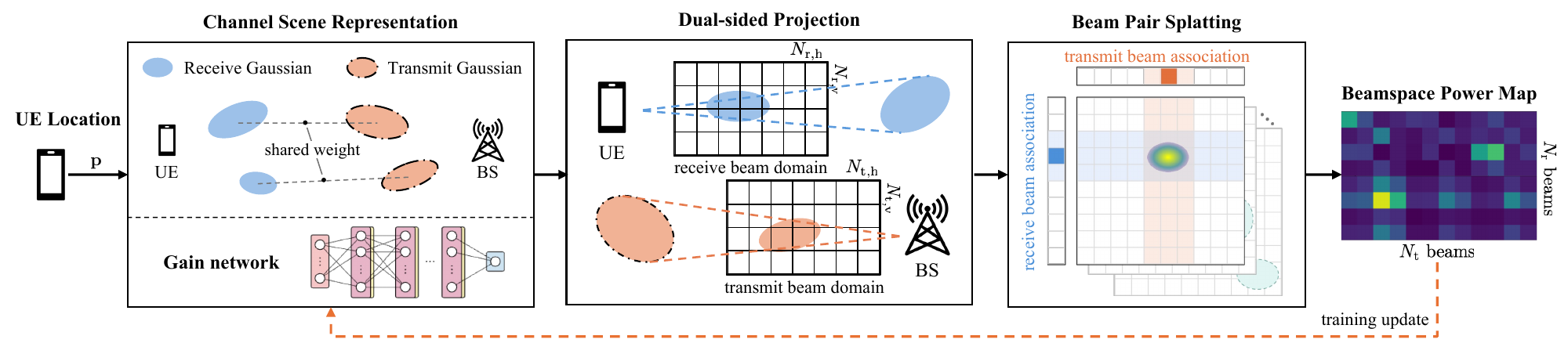}
    \caption{Overall architecture of the MIMO-GS framework. The channel scene is represented by primitives, each pairing a receive Gaussian and a transmit Gaussian placed in the propagation environment. The receive Gaussian is projected from the UE onto the receive beam domain and the transmit Gaussian from the BS onto the transmit beam domain, which determines the receive and transmit DFT beams the primitive falls on. Given a UE location $\mathbf{p}$, each primitive contributes with a static weight shared by the pair multiplied by a gain that the gain network produces from $\mathbf{p}$, and beam pair splatting accumulates the contributions over all primitives into the beamspace power map $\widehat{\mathbf{X}}(\mathbf{p})$. The scene and the gain network are optimized from the rendering error against the observed maps.}
    \label{fig:overall_architecture}
\end{figure*}

3D-GS is an explicit and differentiable scene representation originally developed for real-time radiance field rendering \cite{kerbl:acm:23}. 
Prior to 3D-GS, NeRF \cite{nerf} represented a scene implicitly as a neural network mapping position and viewing direction to density and radiance, so rendering required dense sampling along every ray and a large number of network evaluations, making both training and rendering costly. 
Instead of representing a scene as an implicit neural field through dense volumetric sampling, 3D-GS models the scene as a collection of anisotropic Gaussian primitives. Let $\mathcal{G}=\{g_i\}_{i=1}^{N_{\mathrm{g}}}$ denote a set of $N_{\mathrm{g}}$ 3D Gaussians. Each Gaussian primitive is characterized by its center $\boldsymbol{\mu}_i\in\mathbb{R}^3$, covariance matrix $\boldsymbol{\Sigma}_i\in\mathbb{R}^{3\times 3}$, opacity $o_i$, and a learnable attribute vector $\mathbf{c}_i$. The spatial support of the $i$-th primitive is described by the Gaussian density
\begin{align}
    G_i(\mathbf{x})=\exp\left(-\frac{1}{2}(\mathbf{x}-\boldsymbol{\mu}_i)^{\mathsf{T}}\boldsymbol{\Sigma}_i^{-1}(\mathbf{x}-\boldsymbol{\mu}_i)\right),
\end{align}
where $\mathbf{x}\in\mathbb{R}^3$ denotes a point in the scene. Since $\boldsymbol{\Sigma}_i$ must remain positive semi-definite during optimization, it is commonly parameterized by a rotation matrix $\mathbf{R}_i$ and a scaling matrix $\mathbf{S}_i$ as $\boldsymbol{\Sigma}_i=\mathbf{R}_i\mathbf{S}_i\mathbf{S}_i^{\mathsf{T}}\mathbf{R}_i^{\mathsf{T}}$.

To render an observation, 3D-GS projects the Gaussian primitives onto a two-dimensional observation domain. Under a local affine approximation of the projection operation, the projected covariance of the $i$-th Gaussian can be written as $\boldsymbol{\Sigma}'_i=\mathbf{J}_i\mathbf{W}\boldsymbol{\Sigma}_i\mathbf{W}^{\mathsf{T}}\mathbf{J}_i^{\mathsf{T}}$, where $\mathbf{W}$ denotes the view transformation and $\mathbf{J}_i$ is the Jacobian of the projection evaluated at the Gaussian center. Let $\boldsymbol{\mu}'_i$ denote the projected center. For a sample point $\mathbf{u}$ on the observation plane, the opacity contribution of the projected Gaussian is given by
\begin{align}
    \alpha_i(\mathbf{u})=o_i\exp\left(-\frac{1}{2}(\mathbf{u}-\boldsymbol{\mu}'_i)^{\mathsf{T}}\left(\boldsymbol{\Sigma}'_i\right)^{-1}(\mathbf{u}-\boldsymbol{\mu}'_i)\right).
\end{align}
The projected Gaussians that overlap the same sample point are sorted according to their depth and accumulated through front-to-back $\alpha$-blending. If $\mathcal{S}(\mathbf{u})$ denotes the depth-ordered set of Gaussians contributing to $\mathbf{u}$, the rendered attribute is
\begin{align}
    \widehat{\mathbf{y}}(\mathbf{u}) = \sum_{i\in\mathcal{S}(\mathbf{u})} T_i(\mathbf{u})\alpha_i(\mathbf{u})\mathbf{c}_i,\quad T_i(\mathbf{u})=\prod_{j<i}\left(1-\alpha_j(\mathbf{u})\right),
\end{align}
where $T_i(\mathbf{u})$ denotes the accumulated transmittance before the $i$-th Gaussian. Since all operations in the projection and blending pipeline are differentiable, the Gaussian parameters can be optimized by comparing the rendered outputs with the corresponding observations, and the representation is refined during training by cloning, splitting, or pruning Gaussians. In this work, the rendered quantity is defined in the beamspace power domain rather than as visual appearance such as color, and the Gaussian primitives are used as differentiable latent primitives for representing the MIMO channel scene, rather than as identifiable physical surfaces or scatterers.


\section{MIMO-GS Rendering Framework}
\label{section_mimogs_framework}

Motivated by the 3D-GS framework, we now develop MIMO-GS, which parameterizes the renderer $f_{\Theta}$ with a set of 3D Gaussians and outputs the beam pair power map $\widehat{\mathbf{X}}(\mathbf{p})\in\mathbb{R}_{+}^{N_{\mathrm{r}}\times N_{\mathrm{t}}}$ at a UE location $\mathbf{p}$. As illustrated in Fig.~\ref{fig:overall_architecture}, MIMO-GS consists of three main components: a channel scene representation, a dual-sided projection, and a beam pair splatting procedure. Each primitive pairs a receive Gaussian and a transmit Gaussian. The receive Gaussian is projected from the UE to determine the receive beam association, while the transmit Gaussian is projected from the BS to determine the transmit beam association, and beam pair splatting accumulates the two associations into the map. Since each entry of the target is indexed jointly by a receive and a transmit beam, a single projection from one side cannot form it, which is why the projection is dual-sided.


\subsection{Channel Scene Representation}

In MIMO-GS, we build on the primitive set $\mathcal{G}=\{g_k\}_{k=1}^{N_{\mathrm{g}}}$ introduced in Section~\ref{3dgs} to represent the channel scene. However, since a beam pair power map is jointly indexed by a receive and a transmit DFT beam, each primitive consists of a receive Gaussian with mean $\mathbf{q}^{\mathrm{r}}_k\in\mathbb{R}^{3}$ and covariance $\boldsymbol{\Sigma}_k^{\mathrm{r}}\in\mathbb{R}^{3\times3}$, a transmit Gaussian with mean $\mathbf{q}^{\mathrm{t}}_k\in\mathbb{R}^{3}$ and covariance $\boldsymbol{\Sigma}_k^{\mathrm{t}}\in\mathbb{R}^{3\times3}$, and a shared static weight $\rho_k\in\mathbb{R}_{+}$, that is,
\begin{align}
g_k=\left\{\mathbf{q}^{\mathrm{r}}_k,\boldsymbol{\Sigma}_k^{\mathrm{r}}, \mathbf{q}^{\mathrm{t}}_k,\boldsymbol{\Sigma}_k^{\mathrm{t}},\rho_k\right\}.
\label{5_primitive}
\end{align}

In contrast to Section~\ref{3dgs}, where each primitive is a single Gaussian, a MIMO-GS primitive comprises two Gaussians associated through the shared weight $\rho_k$. The means $\mathbf{q}^{\mathrm{r}}_k$ and $\mathbf{q}^{\mathrm{t}}_k$ determine the beam associations at the UE and the BS, respectively, and serve as effective geometric parameters rather than physical scatterer locations. When they are close or identical, the departure and arrival geometries are confined to the same spatial region, as in a single-bounce interaction, whereas when they are distinct, a mixture of primitives represents reflection, diffraction, diffuse scattering, and multiple-bounce propagation. A primitive whose two Gaussians lie on the segment between the BS and the UE renders the line-of-sight (LoS) beam pair without a scatterer. Each covariance controls the angular spread after projection on its own side, and the shared weight $\rho_k$ represents the static importance of the component before location-dependent modulation. As in Section~\ref{3dgs}, $\boldsymbol{\Sigma}_k^{\mathrm{r}}$ and $\boldsymbol{\Sigma}_k^{\mathrm{t}}$ are parameterized using rotation and scaling variables.

The projection determines which beam pairs a primitive contributes to, but not how much power it contributes. The power depends on the UE location through the visibility of the primitive, which changes abruptly across blockage boundaries, and the projection does not capture it. We therefore factor the gain of the $k$-th primitive into a static weight $\rho_k$, which represents the strength of the primitive in the scene similarly to the opacity in 3D-GS, and a location-dependent factor $d_k(\mathbf{p})$ obtained from a small neural network shared by all primitives, referred to as the gain network. The network takes the primitive and UE locations as input and outputs a single scalar, while the projection and splatting distribute this power over the $N_{\mathrm{r}}\times N_{\mathrm{t}}$ beam pairs, so the network size does not depend on the array sizes and its cost is one forward pass per primitive. It consists of three linear layers with 64 hidden units and ReLU activations, followed by a final layer that outputs the scalar.



Since the power of a primitive can change abruptly across a blockage or visibility boundary, whereas neural networks that take raw coordinates as input have a spectral bias toward slowly varying functions, we apply a positional encoding based on Fourier features to the coordinate inputs. For a given vector $\mathbf{x}\in\mathbb{R}^{3}$, the encoding function is defined as
\begin{align}
    \Gamma(\mathbf{x})
    =
    \left[
    \mathbf{x}^{\mathsf{T}},
    \left\{
    \sin(\omega_\ell \mathbf{x})^{\mathsf{T}},
    \cos(\omega_\ell \mathbf{x})^{\mathsf{T}}
    \right\}_{\ell=0}^{L_\mathrm{pe}-1}
    \right]^{\mathsf{T}},\label{eq:positional_encoding}
\end{align}
where $L_{\mathrm{pe}}$ denotes the number of Fourier frequency bands and $\omega_\ell = 2^\ell \pi$ is the $\ell$-th spatial frequency in radians per coordinate unit. 
Specifically, for the $k$-th primitive and a target UE location $\mathbf{p}$, the network input $\boldsymbol{\psi}_k(\mathbf{p})$ is constructed using their encoded positions and their encoded relative displacement with respect to the mean of the receive Gaussian:
\begin{align}
    \boldsymbol{\psi}_k(\mathbf{p})
    =
    \left[
    \Gamma(\mathbf{q}^{\mathrm{r}}_k)^{\mathsf{T}},
    \Gamma(\mathbf{p})^{\mathsf{T}},
    \Gamma(\mathbf{q}^{\mathrm{r}}_k-\mathbf{p})^{\mathsf{T}}
    \right]^{\mathsf{T}}.
    \label{5_gain_feature}
\end{align}


Given this constructed feature vector, the network $g_{\varphi}(\cdot)$ with parameters $\varphi$ maps $\boldsymbol{\psi}_k(\mathbf{p})$ to a nonnegative dynamic gain factor, and the final gain of the $k$-th primitive is
\begin{align}
    d_k(\mathbf{p})=\sigma_{\mathrm{sp}}\big(g_{\varphi}(\boldsymbol{\psi}_k(\mathbf{p}))\big),\qquad
    \eta_k(\mathbf{p})=\rho_k\, d_k(\mathbf{p}),
    \label{5_dynamic_gain}
\end{align}
where $\sigma_{\mathrm{sp}}(\cdot)$ denotes the softplus function and $d_k(\mathbf{p})$ scales the static weight $\rho_k$ at the UE location $\mathbf{p}$. The distribution of this contribution over the receive and transmit DFT beams is then determined by the projection and splatting operations described next.

\subsection{Dual-Sided Projection onto Beam Domains}

\begin{figure}[t]
    \centering
    \includegraphics[width=0.85\columnwidth]{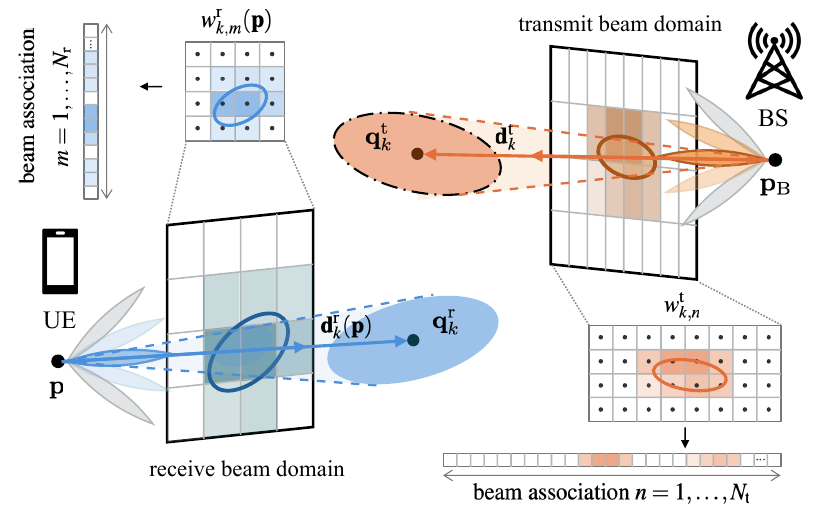}
    \caption{Dual-sided projection of one primitive. The receive Gaussian at $\mathbf{q}^{\mathrm{r}}_k$ is projected from the UE at $\mathbf{p}$ onto the receive beam domain, and the transmit Gaussian at $\mathbf{q}^{\mathrm{t}}_k$ from the BS at $\mathbf{p}_{\mathrm{B}}$ onto the transmit beam domain. The front view of each beam domain shows the beam centers as dots, the projected covariance as an ellipse, and the cells shaded in proportion to the association weight of each beam.}
    \label{fig:how_to_determine}
\end{figure}

The key operation in MIMO-GS, as illustrated in Fig.~\ref{fig:how_to_determine}, is to determine how each primitive contributes to the receive and transmit DFT beams. MIMO-GS performs two geometric projections, one from the UE array for the receive Gaussian and one from the BS array for the transmit Gaussian.

Let $\mathbf{p}_{\mathrm{B}}$ denote the fixed BS location. For a target UE location $\mathbf{p}$, the unit direction from the UE to $\mathbf{q}_k^{\mathrm{r}}$ is
\begin{align}
    \mathbf{d}^{\mathrm{r}}_k(\mathbf{p})
    =
    \frac{\mathbf{q}^{\mathrm{r}}_k-\mathbf{p}}
    {\|\mathbf{q}^{\mathrm{r}}_k-\mathbf{p}\|_2}.
    \label{5_rx_direction}
\end{align}
Similarly, the unit direction from the BS to the center of the transmit Gaussian is
\begin{align}
    \mathbf{d}^{\mathrm{t}}_k
    =
    \frac{\mathbf{q}^{\mathrm{t}}_k-\mathbf{p}_{\mathrm{B}}}
    {\|\mathbf{q}^{\mathrm{t}}_k-\mathbf{p}_{\mathrm{B}}\|_2}.
    \label{5_tx_direction}
\end{align}
With $\mathbf{q}^{\mathrm{r}}_k$, $\mathbf{q}^{\mathrm{t}}_k$, $\mathbf{p}$, and $\mathbf{p}_{\mathrm{B}}$ expressed in a global coordinate frame, let $\mathbf{Q}_{\mathrm{r}},\mathbf{Q}_{\mathrm{t}}\in\mathbb{R}^{3\times3}$ denote the rotation matrices whose columns are the axes of the UE and BS array frames expressed in this global frame, so that $\mathbf{Q}_{\mathrm{r}}^{\mathsf{T}}$ and $\mathbf{Q}_{\mathrm{t}}^{\mathsf{T}}$ map a global direction into the respective array frame. The array orientations are assumed fixed and known throughout the scene. The receive and transmit beam coordinates of the $k$-th primitive are obtained by applying the corresponding rotation matrix to express each direction vector relative to the array, and extracting the two direction cosines that index the UPA beamspace:
\begin{align}
    \mathbf{u}^{\mathrm{r}}_k(\mathbf{p})
    =
    \pi_{\mathrm{r}}
    \left(
    \mathbf{q}^{\mathrm{r}}_k;\mathbf{p}
    \right)
    =
    \mathbf{E}\,\mathbf{Q}_{\mathrm{r}}^{\mathsf{T}}\,\mathbf{d}^{\mathrm{r}}_k(\mathbf{p})
    \in
    \mathbb{R}^{2},
    \label{5_rx_beam_coordinate}
\end{align}
\begin{align}
    \mathbf{u}^{\mathrm{t}}_k
    =
    \pi_{\mathrm{t}}
    \left(
    \mathbf{q}^{\mathrm{t}}_k;\mathbf{p}_{\mathrm B}
    \right)
    =
    \mathbf{E}\,\mathbf{Q}_{\mathrm{t}}^{\mathsf{T}}\,\mathbf{d}^{\mathrm{t}}_k
    \in
    \mathbb{R}^{2},
    \label{5_tx_beam_coordinate}
\end{align}
where $\mathbf{E}\in\mathbb{R}^{2\times3}$ is a selection matrix that extracts the two coordinates used for the UPA beam coordinate domain from a unit vector expressed relative to the array, with its specific form depending on the array’s geometric convention. The transmit mapping $\pi_{\mathrm{t}}(\cdot)$ follows an identical structure with respect to the BS array, using $\mathbf{Q}_{\mathrm{t}}$ and $\mathbf{p}_{\mathrm{B}}$.

The spatial spread of each Gaussian must also be reflected in the beam domain, since a large or elongated Gaussian covers several beams whereas a compact one affects few. MIMO-GS therefore projects the covariance of each side into its own beam coordinate domain, using the local affine approximation around the Gaussian mean that 3D-GS applies to its image plane. The Jacobian of $\pi_{\mathrm{r}}$ evaluated at $\mathbf{q}^{\mathrm{r}}_k$ is obtained by direct differentiation as
\begin{align}
    \mathbf{J}^{\mathrm{r}}_k(\mathbf{p}) = \frac{\partial \pi_{\mathrm{r}}(\mathbf{x};\mathbf{p})}{\partial \mathbf{x}}\Big|_{\mathbf{x}=\mathbf{q}^{\mathrm{r}}_k} = \frac{\mathbf{E}\,\mathbf{Q}_{\mathrm{r}}^{\mathsf{T}}\big(\mathbf{I}_3 - \mathbf{d}^{\mathrm{r}}_k(\mathbf{d}^{\mathrm{r}}_k)^{\mathsf{T}}\big)}{\|\mathbf{q}^{\mathrm{r}}_k-\mathbf{p}\|_2}, \label{5_rx_jacobian}
\end{align}
which captures two effects: the orthogonal projection $\mathbf{I}_3 - \mathbf{d}^{\mathrm{r}}_k(\mathbf{p})\,\mathbf{d}^{\mathrm{r}}_k(\mathbf{p})^{\mathsf{T}}$ onto the tangent plane perpendicular to $\mathbf{d}^{\mathrm{r}}_k(\mathbf{p})$, and the factor $1/\|\mathbf{q}^{\mathrm{r}}_k - \mathbf{p}\|_2$ which reflects that the angular size of a projected Gaussian decreases with distance. The projected covariance at the receive array is then
\begin{align}
    \mathbf{C}^{\mathrm{r}}_k(\mathbf{p})
    =
    \mathbf{J}^{\mathrm{r}}_k(\mathbf{p})
    \boldsymbol{\Sigma}_k^{\mathrm{r}}
    \left(
    \mathbf{J}^{\mathrm{r}}_k(\mathbf{p})
    \right)^{\mathsf{T}}.
    \label{5_rx_projected_covariance}
\end{align}
Similarly, the projected covariance at the transmit array is $\mathbf{C}^{\mathrm{t}}_k=\mathbf{J}^{\mathrm{t}}_k\boldsymbol{\Sigma}_k^{\mathrm{t}}(\mathbf{J}^{\mathrm{t}}_k)^{\mathsf{T}}$ with
\begin{align}
    \mathbf{J}^{\mathrm{t}}_k = \frac{\partial \pi_{\mathrm{t}}(\mathbf{x};\mathbf{p}_{\mathrm{B}})}{\partial \mathbf{x}}\Big|_{\mathbf{x}=\mathbf{q}^{\mathrm{t}}_k} = \frac{\mathbf{E}\,\mathbf{Q}_{\mathrm{t}}^{\mathsf{T}}\big(\mathbf{I}_3 - \mathbf{d}^{\mathrm{t}}_k(\mathbf{d}^{\mathrm{t}}_k)^{\mathsf{T}}\big)}{\|\mathbf{q}^{\mathrm{t}}_k - \mathbf{p}_{\mathrm{B}}\|_2}. \label{5_tx_jacobian}
\end{align}

Now let $\{\mathbf{b}^{\mathrm{r}}_m\}_{m=1}^{N_{\mathrm{r}}}$ and $\{\mathbf{b}^{\mathrm{t}}_n\}_{n=1}^{N_{\mathrm{t}}}$ denote the receive and transmit DFT beam centers in this two-dimensional beam coordinate domain. The association weight between the receive Gaussian of the $k$-th primitive and the $m$-th receive beam is computed by evaluating the projected Gaussian density at $\mathbf{b}^{\mathrm{r}}_m$:
\begin{align}
    \widetilde{w}^{\mathrm{r}}_{k,m}(\mathbf{p})
    =
    \exp
    \left(
    -\frac{1}{2}
    \left(
    \mathbf{b}^{\mathrm{r}}_m
    -
    \mathbf{u}^{\mathrm{r}}_k(\mathbf{p})
    \right)^{\mathsf{T}}
    \left(
    \mathbf{C}^{\mathrm{r}}_k(\mathbf{p})
    \right)^{-1}
    \left(
    \mathbf{b}^{\mathrm{r}}_m
    -
    \mathbf{u}^{\mathrm{r}}_k(\mathbf{p})
    \right)
    \right).
    \label{5_rx_raw_weight}
\end{align}
Similarly, the transmit beam association is computed in the same manner:
\begin{align}
    \widetilde{w}^{\mathrm{t}}_{k,n}
    =
    \exp
    \left(
    -\frac{1}{2}
    \left(
    \mathbf{b}^{\mathrm{t}}_n
    -
    \mathbf{u}^{\mathrm{t}}_k
    \right)^{\mathsf{T}}
    \left(
    \mathbf{C}^{\mathrm{t}}_k
    \right)^{-1}
    \left(
    \mathbf{b}^{\mathrm{t}}_n
    -
    \mathbf{u}^{\mathrm{t}}_k
    \right)
    \right).
    \label{5_tx_raw_weight}
\end{align}
For the Gaussian beam association weights, the projected covariances $\mathbf{C}^{\mathrm{r}}_k(\mathbf{p})$ and $\mathbf{C}^{\mathrm{t}}_k$ are regularized with a small positive diagonal term to ensure invertibility and to maintain a minimum spread in the beam coordinate domain. The weights are then normalized per Gaussian over their respective beam indices as
\begin{align}
    w^{\mathrm{r}}_{k,m}(\mathbf{p})
    =
    \frac{
    \widetilde{w}^{\mathrm{r}}_{k,m}(\mathbf{p})
    }{
    \sum_{m'=1}^{N_{\mathrm r}}
    \widetilde{w}^{\mathrm{r}}_{k,m'}(\mathbf{p})
    },\quad
    w^{\mathrm{t}}_{k,n}
    =
    \frac{
    \widetilde{w}^{\mathrm{t}}_{k,n}
    }{
    \sum_{n'=1}^{N_{\mathrm t}}
    \widetilde{w}^{\mathrm{t}}_{k,n'}
    }.
\end{align}
Since the two associations are formed independently, $\mathbf{q}^{\mathrm{r}}_k$ and $\mathbf{q}^{\mathrm{t}}_k$ parameterize the angular support of an effective beam pair component rather than a single physical interaction point, and propagation involving several interactions is represented by the superposition of primitives.

\subsection{Beam Pair Splatting for Rendering}

After obtaining the receive and transmit beam association weights, MIMO-GS renders each beam pair entry by accumulating Gaussian contributions. For the beam pair indexed by $(m,n)$, the power contribution of the $k$-th primitive is defined as
\begin{align}
    c_{k,m,n}(\mathbf{p})
    =
    \eta_k(\mathbf{p})
    w^{\mathrm{r}}_{k,m}(\mathbf{p})
    w^{\mathrm{t}}_{k,n},
    \label{5_Gaussian_beam_pair_contribution}
\end{align}
where $\eta_k(\mathbf{p})$ represents the location-dependent gain of the $k$-th primitive at the target UE location, while $w^{\mathrm{r}}_{k,m}(\mathbf{p})$ and $w^{\mathrm{t}}_{k,n}$ determine the allocation of this weight to the $m$-th receive and $n$-th transmit DFT beams, respectively. The product $w^{\mathrm r}_{k,m}(\mathbf p)w^{\mathrm t}_{k,n}$ defines a separable allocation rule over the receive and transmit beam indices. Summing over all primitives gives the rendered power of the $(m,n)$-th beam pair:
\begin{align}
    \widehat{X}_{m,n}(\mathbf{p})
    =
    \sum_{k=1}^{N_{\mathrm{g}}}
    c_{k,m,n}(\mathbf{p})
    =
    \sum_{k=1}^{N_{\mathrm{g}}}
    \eta_k(\mathbf{p})
    w^{\mathrm{r}}_{k,m}(\mathbf{p})
    w^{\mathrm{t}}_{k,n}.
    \label{5_beam_pair_rendering}
\end{align}
Stacking all entries gives the rendered beam pair power map
\begin{align}
    \widehat{\mathbf{X}}(\mathbf{p})
    =
    f_{\Theta}(\mathbf{p})
    \in
    \mathbb{R}_{+}^{N_{\mathrm{r}}\times N_{\mathrm{t}}}.
\end{align}
Equation~\eqref{5_beam_pair_rendering} is the main rendering equation of MIMO-GS, replacing splatting over image pixels with splatting over receive and transmit DFT beam pairs.

The aggregation in \eqref{5_beam_pair_rendering} employs direct summation of Gaussian contributions, in contrast to the depth-ordered $\alpha$-blending used in optical 3D-GS and prior RF radiance field methods. The $\alpha$-blending based on depth ordering is not appropriate for this target because the rendered quantity is not optical radiance accumulated along a camera ray. In optical rendering, the transmittance $T_i = \prod_{j<i}\left(1-\alpha_j\right)$ encodes that an opaque surface occludes whatever lies behind it along the viewing ray. A beam pair power map, however, does not provide a universal depth-ordering of Gaussians. Therefore, MIMO-GS uses direct nonnegative summation as an effective aggregation rule over the beam pair power domain. Visibility and blockage effects are absorbed into the learned location-dependent gain $\eta_k(\mathbf{p})$, rather than represented through optical $\alpha$-transmittance.

This aggregation rule in \eqref{5_beam_pair_rendering} is consistent with the multipath structure of the target. In a conventional multipath channel, each path contributes a localized response around its associated transmit and receive angular coordinates in the beamspace domain, so that the beam pair power becomes
\begin{align}
X_{m,n}(\mathbf{p})
= \mathbb{E}\big[\big|\textstyle\sum_{\ell \in \mathcal{R}} \beta_\ell
[\mathbf{A}_{\mathrm{r}}^{\mathsf{H}}\mathbf{a}_{\mathrm{r}}(\boldsymbol{\Omega}^{\mathrm{r}}_\ell)]_m
[\mathbf{a}_{\mathrm{t}}^{\mathsf{H}}(\boldsymbol{\Omega}^{\mathrm{t}}_\ell)\mathbf{A}_{\mathrm{t}}]_n\big|^2\big],
\end{align}
where the dependence on $\mathbf{p}$ is omitted inside the sum.
Since the rendering target is the beam pair power map, MIMO-GS models the aggregate power domain structure induced by the coherent multipath response, rather than resolving individual complex path coefficients. Motivated by the localized support of multipath components in beamspace, each projected primitive provides a learnable nonnegative contribution over the beam pair grid, and the rendered map is obtained by accumulating these contributions.


\subsection{Training and Inference}

The learnable parameters of MIMO-GS are collected as
\begin{align}
    \Theta
    =
    \left\{
    \mathbf{q}^{\mathrm{r}}_k,
    \boldsymbol{\Sigma}_k^{\mathrm{r}},
    \mathbf{q}^{\mathrm{t}}_k,
    \boldsymbol{\Sigma}_k^{\mathrm{t}},
    \rho_k
    \right\}_{k=1}^{N_{\mathrm{g}}}
    \cup
    \{\varphi\},
\end{align}
where $\varphi$ denotes the parameters of the gain network. Given the observation set $\mathcal{D}=\{(\mathbf{p}_i,\mathbf{X}_i)\}_{i=1}^{N}$, MIMO-GS renders $\widehat{\mathbf{X}}(\mathbf{p}_i)$ for each observed UE location and updates $\Theta$ by minimizing the reconstruction loss defined in the problem formulation. Since the gain prediction, Gaussian projection, beam association, and beam pair splatting operations are differentiable, the Gaussian parameters and the gain network parameters can be optimized jointly.

MIMO-GS periodically applies adaptive density control during training, adjusting the number of primitives through densification and pruning. Densification targets primitives with large positional gradients, which indicate that moving them reduces the reconstruction loss. A primitive of small spatial extent, taken as the larger of the maximum scales of $\boldsymbol{\Sigma}_k^{\mathrm{r}}$ and $\boldsymbol{\Sigma}_k^{\mathrm{t}}$, is cloned and slightly shifted, whereas one that overlaps many DFT beams is split into two smaller primitives to resolve distinct multipath clusters. Pruning removes from $\mathcal{G}$ the primitives whose rendered contribution falls below a threshold $\rho_\mathrm{min}$, so that the rendering capacity stays on the dominant beam pairs rather than on empty regions.

After training, the renderer can be evaluated at an arbitrary UE location $\mathbf{p}^\star$. The trained model first computes the primitive gains for $\mathbf{p}^\star$, then projects the two Gaussians of each primitive onto their respective beam domains, and finally applies the beam pair splatting rule in \eqref{5_beam_pair_rendering}. The resulting output $\widehat{\mathbf{X}}(\mathbf{p}^\star)$ is the rendered beam pair power map at an arbitrary UE location $\mathbf{p}^\star$ within the considered environment.

\begin{remark} \normalfont
Prior RF adaptations of 3D-GS build on a link with a single antenna at one end, so their rendered domain is an angular grid observed from one side (Table~\ref{table_related_works}); in RF-3DGS \cite{rf3dgs} the angle of departure is an attribute attached to the receive map, and in BiWGS \cite{biwgs} the bidirectionality resides in the scattering function of each primitive while the output is still observed from the receiver. The beam pair power map cannot be assembled from such one-sided outputs, which identify the strong beams on each side but not which receive beam is paired with which transmit beam. MIMO-GS instead projects each primitive twice, and every output entry is indexed jointly by the two projections, so the coupling between departure and arrival directions is carried by the primitive itself. The rendered domain is the 2D-DFT beam grid of each UPA rather than an angular hemisphere, so each entry corresponds to a DFT beam resource used for beam management, and the grid size is set by the codebook rather than by a chosen angular resolution.
\end{remark}

\section{Beam Management with Rendered Maps}
\label{section_beam_management}

Beam management is a primary application of the rendered map. In 5G NR, the BS sweeps beamformed CSI-RS resources and the UE measures and reports them \cite{3gpp38214}; as the array grows, the beams narrow and the sweep lengthens, and the overhead grows further with the number of users served \cite{giordani2019bm}.

\subsection{Multiuser Transmission in the Beam Domain} 

Prior side information schemes predict a small number of BS beam candidates from the UE position or sensor data \cite{va2019}, or learn a fixed set of measurement beams for the deployment \cite{heng2022}. When several users are served at the same time on distinct beams, however, the power on a user's own beam alone does not determine its rate. The beams of the other scheduled users also reach it, and the interference they contribute takes a different value for every user and every beam. Scheduling therefore requires the power of every beam at every user, not the index of the best one. MIMO-GS renders exactly this quantity conditioned on the UE location, as defined in \eqref{2_power_map}.


The transmission follows the beam division multiple access (BDMA) framework \cite{sun2015bdma}, in which the BS transmits on DFT beams without instantaneous CSI at the transmitter, each transmit beam carries one data stream and is assigned to at most one user, and the users are separated by the beams they occupy. The BS then needs only the statistical CSI of each user in the beam domain, which is the beam pair power map of \eqref{2_power_map}. Acquiring it conventionally requires channel measurements. Each user is sounded on every beam over many coherence intervals, so the pilot cost grows with the codebook size. A user at initial access has no measurement history and cannot be scheduled until it is sounded. MIMO-GS instead renders the same quantity from the reported position, so once the scene is trained from channel observations collected in the deployment, the per user sounding is replaced by a rendering pass while the transmission and the scheduling that operates on it are unchanged. Each stream is assigned equal power, so that the gain comes from the choice of users and beams rather than from power allocation.

Let $U$ candidate UEs at positions $\mathbf{p}_{1},\ldots,\mathbf{p}_{U}$ be served by a scheduled set $\mathcal{S}$ of at most $L$ users. Scheduled user $k$ is assigned a transmit beam $n_{k}$, distinct across users, and receives through a single receive beam $m_{k}$. With $\mathbf{f}_{\mathrm{t},n}$ and $\mathbf{w}_{\mathrm{r},m}$ the $n$-th and $m$-th columns of $\mathbf{A}_{\mathrm{t}}$ and $\mathbf{A}_{\mathrm{r}}$, the BS transmits $\mathbf{x} = \sum_{j\in\mathcal{S}} \mathbf{f}_{\mathrm{t},n_{j}} s_{j},$ where $s_{j}\sim\mathcal{CN}\left(0, P/|\mathcal{S}|\right)$ denotes the data stream of user $j$ and $P$ the total transmit power. User $k$ observes after receive beamforming
\begin{align}
    \label{6_mu_rx}
    y_{k} &= \mathbf{w}_{\mathrm{r},m_{k}}^{\mathsf{H}}\left(\mathbf{H}(\mathbf{p}_{k})\,\mathbf{x} + \mathbf{z}_{k}\right) \notag\\
    &= [\mathbf{H}_{\mathrm{b}}(\mathbf{p}_{k})]_{m_{k},n_{k}} s_{k}
    + \sum_{j\in\mathcal{S},\,j\neq k} [\mathbf{H}_{\mathrm{b}}(\mathbf{p}_{k})]_{m_{k},n_{j}} s_{j}
    + \tilde{z}_{k},
\end{align}
where $\mathbf{z}_{k}\sim\mathcal{CN}\left(\mathbf{0},\sigma_{z}^{2}\mathbf{I}_{N_{\mathrm{r}}}\right)$ is the receive noise with power $\sigma_{z}^{2}$, $\tilde{z}_{k}=\mathbf{w}_{\mathrm{r},m_{k}}^{\mathsf{H}}\mathbf{z}_{k}$, and the second term is the interference from the beams of the other scheduled users. Since the columns of $\mathbf{A}_{\mathrm{t}}$ and $\mathbf{A}_{\mathrm{r}}$ are orthonormal, the transmit power constraint is preserved and the projection onto the receive beam preserves the noise variance $\sigma_{z}^{2}$. Treating the interference as noise, the SINR of user $k$ is
\begin{align}
    \label{6_mu_sinr}
    \mathrm{SINR}_{k}
    =
    \frac{\big|[\mathbf{H}_{\mathrm{b}}(\mathbf{p}_{k})]_{m_{k},n_{k}}\big|^{2}}
    {\sum_{j\in\mathcal{S},\,j\neq k}\big|[\mathbf{H}_{\mathrm{b}}(\mathbf{p}_{k})]_{m_{k},n_{j}}\big|^{2}+\sigma_{z}^{2}|\mathcal{S}|/P}.
\end{align}

Once the users and beams are fixed, the BS configures CSI-RS on the scheduled beams, from which each UE obtains the instantaneous channel over those resources and sets its receive beam. Let $T_{\mathrm{B}}$ denote the beam management interval between beam updates, and $\tau_{\mathrm{RS}}$ the effective overhead per configured transmit CSI-RS resource, including the UE measurements it triggers. Reporting and control signaling overheads, which are strongly implementation dependent and common to the schemes under comparison, fall outside the scope of this paper. Only the $|\mathcal{S}|$ beams in use are configured, so the overhead is $|\mathcal{S}|\tau_{\mathrm{RS}}$ against $N_{\mathrm{t}}\tau_{\mathrm{RS}}$ for a sweep of the full codebook. Assuming $|\mathcal{S}|\tau_{\mathrm{RS}}<T_{\mathrm{B}}$, the net sum rate is
\begin{align}
    \label{6_mu_rate}
    R_{\mathrm{sum}}
    =
    \left(1-\frac{|\mathcal{S}|\,\tau_{\mathrm{RS}}}{T_{\mathrm{B}}}\right)
    \sum_{k\in\mathcal{S}}
    \mathbb{E}_{\mathbf{H}\mid\mathbf{p}_{k}}\left[\log_{2}\left(1+\mathrm{SINR}_{k}\right)\right],
\end{align}
where the expectation is over the channel realizations at $\mathbf{p}_{k}$.

\subsection{Scheduling on the Rendered Map}

The scheduler works with the expected powers provided by the power map, rather than the instantaneous channel. These entries are compared with the noise power and with one another, so they are needed on a common scale. The path gain that sets this scale is estimated from the reported position, so the scheduling needs no measurement of the user candidates. Replacing the instantaneous powers in \eqref{6_mu_sinr} by the rendered entries gives
\begin{align}
    \label{6_mu_metric}
    \widehat{\gamma}_{k}(\mathcal{S},\{n_{j}\},m)
    =
    \frac{[\widehat{\mathbf{X}}(\mathbf{p}_{k})]_{m,n_{k}}}
    {\sum_{j\in\mathcal{S},\,j\neq k}[\widehat{\mathbf{X}}(\mathbf{p}_{k})]_{m,n_{j}}+\sigma_{z}^{2}|\mathcal{S}|/P},
\end{align}
and the scheduling objective
\begin{align}
    \label{6_mu_objective}
    F(\mathcal{S},\{n_{j}\})=\sum_{k\in\mathcal{S}}\max_{m}\log_{2}\left(1+\widehat{\gamma}_{k}(\mathcal{S},\{n_{j}\},m)\right).
\end{align}
The maximization over $m$ predicts the receive beam that the UE will later select from its measurements. Substituting expected powers into the SINR approximates the ergodic rate, and the approximation is used only to choose the users and beams.

\begin{algorithm}[t]
\caption{Multiuser beam scheduling on the rendered map}
\label{alg_mu_sched}
\KwIn{Rendered maps $\{\widehat{\mathbf{X}}(\mathbf{p}_{k})\}_{k=1}^{U}$, maximum number of users $L$}
\KwOut{Scheduled set $\mathcal{S}$, transmit beams $\{n_{k}\}_{k\in\mathcal{S}}$}
$\mathcal{S}\gets\emptyset$, $\mathcal{B}\gets\emptyset$ \tcp*{$\mathcal{B}$: beams in use}
\Repeat{$|\mathcal{S}|=L$}{
    \ForAll{$k\notin\mathcal{S}$ \textnormal{and} $n\notin\mathcal{B}$}{
        $\mathcal{S}'\gets\mathcal{S}\cup\{k\}$, $n_{k}\gets n$\;
        $m_{u}\gets\arg\max_{m}\widehat{\gamma}_{u}(\mathcal{S}',\{n_{j}\},m)$, $\forall u\in\mathcal{S}'$\;
        $\Delta(k,n)\gets F(\mathcal{S}',\{n_{j}\})-F(\mathcal{S},\{n_{j}\})$\;
    }
    $(k^{\star},n^{\star})\gets\arg\max_{k,n}\Delta(k,n)$\;
    \lIf{$\Delta(k^{\star},n^{\star})\le 0$}{\textbf{break}}
    $\mathcal{S}\gets\mathcal{S}\cup\{k^{\star}\}$, $\mathcal{B}\gets\mathcal{B}\cup\{n^{\star}\}$\;
}
\Repeat{no move increases $F$}{
    \ForAll{$k\in\mathcal{S}$ \textnormal{and} $n\notin\mathcal{B}$}{
        \If{$F(\mathcal{S},\{n_{j}\}|_{n_{k}=n})>F(\mathcal{S},\{n_{j}\})$}{
            $\mathcal{B}\gets\left(\mathcal{B}\setminus\{n_{k}\}\right)\cup\{n\}$, $n_{k}\gets n$\;
        }
    }
}
\Return $\mathcal{S}$, $\{n_{k}\}$
\end{algorithm}

Users and beams are allocated one pair at a time. Starting from an empty set, each step adds the pair of a candidate user and a free transmit beam that increases $F$ the most, with the beams of the users already scheduled held fixed and the receive beams of all of them recomputed. The step is repeated until no pair increases $F$ or $|\mathcal{S}|$ reaches $L$, so $L$ is an upper bound and fewer users may be scheduled. Assigned beams are then moved to free beams that increase $F$ until no move is found, since a beam chosen early need not remain the best once later users are present. Since $F$ accounts for the interference, a user may be placed on a beam other than its strongest, or left unscheduled when it would interfere with several others. Algorithm~\ref{alg_mu_sched} summarizes the procedure. Once the assignment is complete, the BS configures CSI-RS on the scheduled beams, and the rates in \eqref{6_mu_rate} are obtained from the measured channels over those resources.

Side information has been used to reduce the beam management overhead, for example by learning a classifier from the UE position to the beams most likely to be optimal \cite{rezaie2022} or by predicting the beams from camera images captured at the BS \cite{alrabeiah2020vision, ahmed23deepsense}. Common to these approaches is that the learned model maps the side information directly to a beam decision, with the strongest beam of a measured sweep as the training label. Its output is one beam index per user and cannot express the interference contributed by another user's beam, so the scheduling metric \eqref{6_mu_metric} is not defined on it. The decision is also tied to the codebook and the selection criterion under which the model was trained, so a change in either requires retraining. MIMO-GS takes the same UE position as input but learns the beam pair powers rather than the decision. A different scheduling rule is handled by re-evaluating \eqref{6_mu_metric} on the same rendered map, and a different codebook only relocates the beam centers at which the scene is rendered. A vision-based predictor is further constrained by its sensing modality. It requires the UE to appear in the camera field of view, and the optical scene does not expose the reflected paths that carry non-line-of-sight (NLoS) links, so its applicability centers on LoS-dominated conditions. The scene in MIMO-GS is instead learned from channel observations, so reflected and scattered components are encoded in the primitives regardless of optical visibility, and the rendering does not presume a LoS link between the BS and the UE.

\section{Numerical Results}

\subsection{Simulation Settings}

\begin{figure}[t]
 \renewcommand{\figurename}{Fig.}
    \centering
    \subfigure[]{\includegraphics[width=0.7\columnwidth]{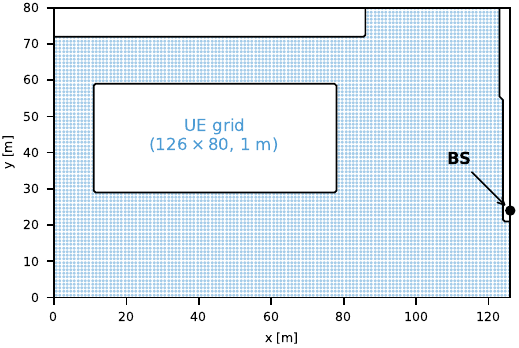}\label{fig:scenario_deepmimo}}\\
    \subfigure[]{\includegraphics[width=0.7\columnwidth]{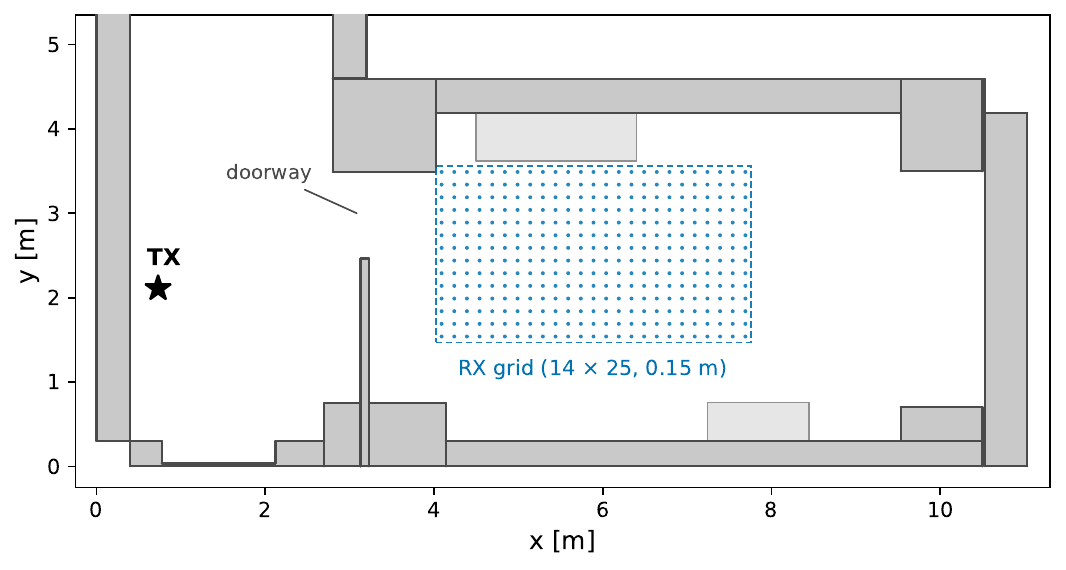}\label{fig:scenario_measured}}
    \caption{Evaluation scenarios. (a) Simulated urban scenario from the ASU campus scenario of DeepMIMO, with the BS location and the UE grid of the channel scene. (b) Measured indoor scenario with the transmitter in the corridor and the receiver grid inside the room.}
    \label{fig:scenarios}
\end{figure}


The DeepMIMO dataset represents an outdoor urban scenario at $3.5$~GHz with one fixed BS \cite{deepmimo}. The BS and the UE are equipped with $16\times16$ and $4\times4$ UPAs, yielding $N_\mathrm{t}=256$ and $N_\mathrm{r}=16$ DFT beams. Both UPAs lie in the $yz$ plane with the boresight along $-x$ toward the UE region, so that in~\eqref{5_rx_beam_coordinate} $\mathbf{Q}_{\mathrm{r}}=\mathbf{Q}_{\mathrm{t}}=\mathbf{I}_3$ and $\mathbf{E}$ extracts the $y$ and $z$ components of the unit direction. The ground truth maps are computed from the path powers and angles of the scenario, with the path contributions summed in power rather than in amplitude. The channel scene covers a $126\times80$~m region in front of the BS, as shown in Fig.~\ref{fig:scenarios}(a), and the $7{,}240$ UE locations lie on a $1$~m grid partitioned into $5{,}792$ training and $1{,}448$ test locations.

The scene is initialized with $25{,}000$ primitives as defined in~\eqref{5_primitive}, i.e., $50{,}000$ Gaussians, and the parameters are optimized with Adam on an NVIDIA RTX 4080 GPU. The maps are normalized by the largest entry over the training set, so that the strongest beam pair in the scene has unit power. With the normalized mean squared error (NMSE) $\varepsilon(\mathbf{A},\mathbf{B})=\|\mathbf{A}-\mathbf{B}\|_{\mathrm{F}}^2/\|\mathbf{B}\|_{\mathrm{F}}^2$ and $\widehat{\mathbf{X}}_\mathrm{n}$ and $\mathbf{X}_\mathrm{n}$ the rendered and observed maps each normalized by their own peak entry, the renderer is trained with
\begin{align}
    \mathcal{L}=w_1\,\varepsilon\big(\widehat{\mathbf{X}},\mathbf{X}_\mathrm{n}\big)+
    w_2\,\varepsilon\big(\widehat{\mathbf{X}}_\mathrm{n},\mathbf{X}_\mathrm{n}\big)+
    w_3\,\varepsilon_{\mathcal{K}}\big(\widehat{\mathbf{X}}_\mathrm{n},\mathbf{X}_\mathrm{n}\big),
\end{align}
where $\varepsilon_{\mathcal{K}}$ restricts the error to the set $\mathcal{K}$ of the strongest entries of $\mathbf{X}_\mathrm{n}$. The first term fixes the peak of the rendered map at one, and the other two constrain the distribution of power over the beam pairs and the strongest beam pairs, respectively. We set $(w_1, w_2, w_3) = (0.4, 0.4, 0.2)$ and take $\mathcal{K}$ as the strongest $1/16$ of the entries, which carry on average over 90\% of the map power across the training locations, so that the third term concentrates on the dominant beam pairs. The rendering accuracy is measured by $\varepsilon(\widehat{\mathbf{X}}_\mathrm{n},\mathbf{X}_\mathrm{n})$, which compares the distribution of power over the beam pairs, averaged over the test locations in the dB domain. The Top-$K$ accuracy is the fraction of the $K$ strongest ground truth beam pairs found among the $K$ strongest rendered entries, and $C_4$ is the ground truth power carried by the four strongest rendered beam pairs divided by that carried by the four strongest ground truth pairs.

The rendered maps are compared against four baselines. Ray tracing is performed with Sionna RT \cite{hoydis2023sionna} for the same scene under the parameter setting of the DeepMIMO scenario. A nearest neighbor outputs the map of the closest training location. An MLP~\cite{tancik2020} and a CNN decoder~\cite{radford2016}, hereafter CNN, learn the mapping from the UE position to the map directly with about the same number of parameters as MIMO-GS and the same loss and training data. The RF rendering methods in Table~\ref{table_related_works} do not produce the joint map and are therefore not compared numerically.

\subsection{Rendering Accuracy and Cost}

\begin{figure}[t]
 \renewcommand{\figurename}{Fig.}
    \centering
    \includegraphics[width=0.8\columnwidth]{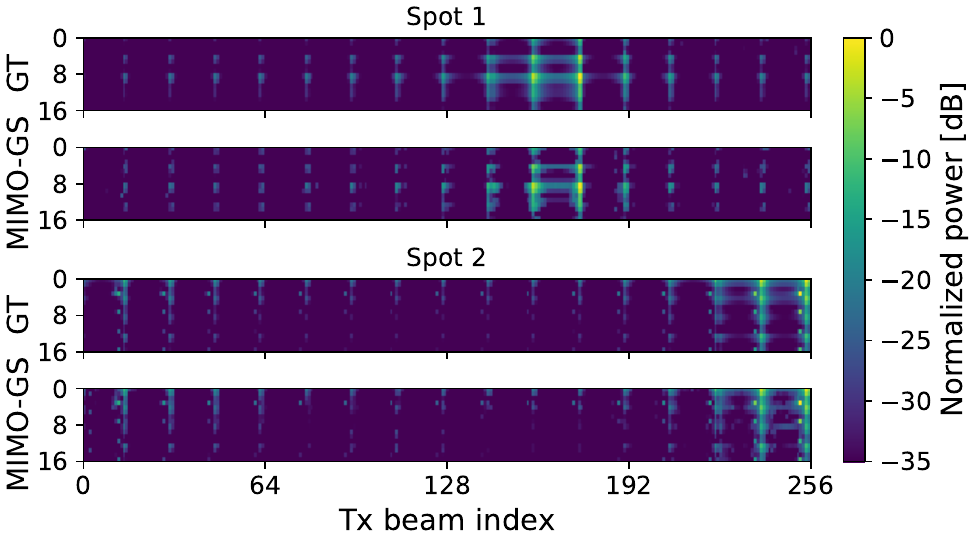}
    \caption{Ground truth and rendered beam pair power maps at two test locations on the DeepMIMO dataset.}
    \label{Fig:samples_deepmimo}
\end{figure}


\begin{figure}[t]
 \renewcommand{\figurename}{Fig.}
    \centering
    \subfigure[]{\includegraphics[width=0.8\columnwidth]{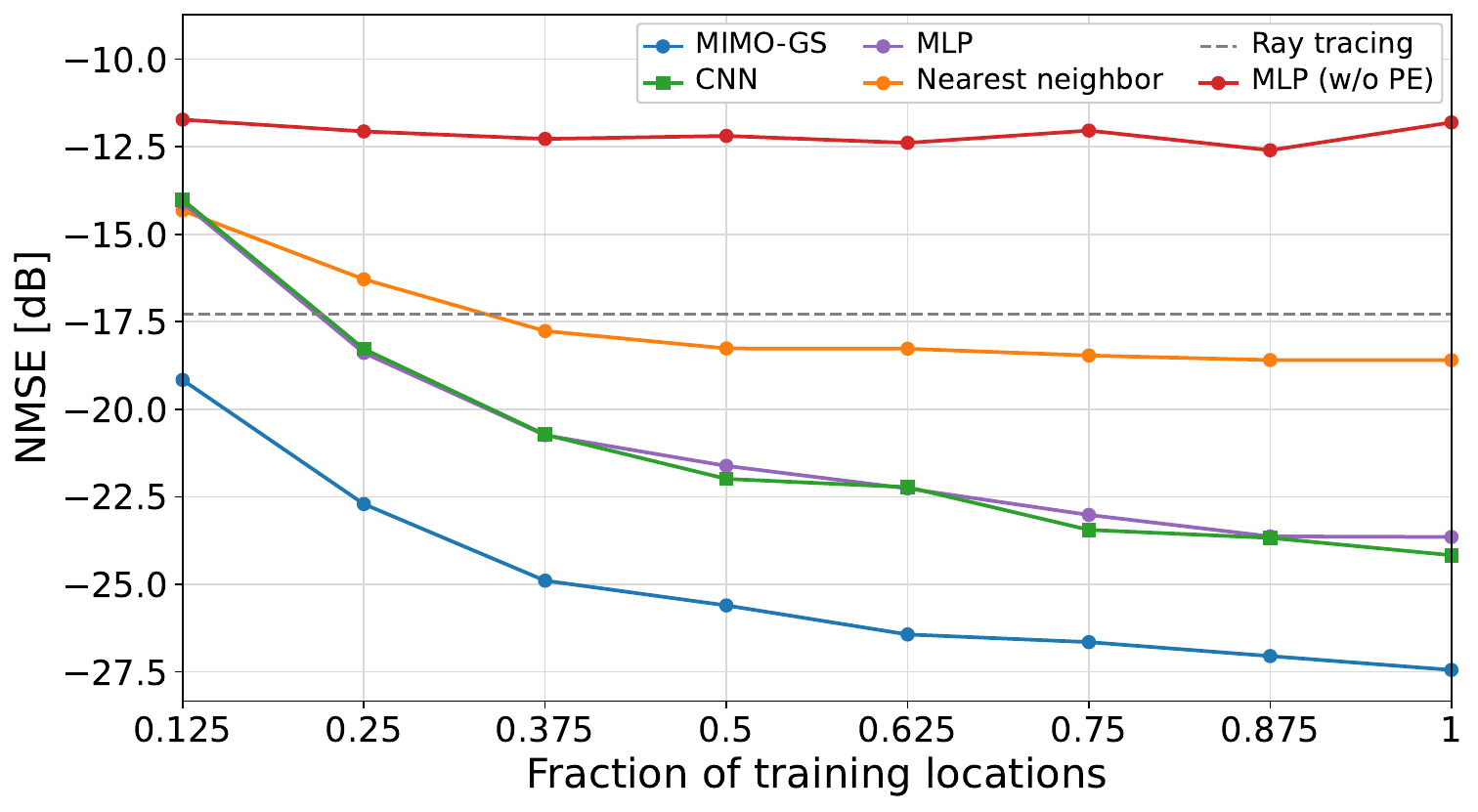}\label{Fig:nmse_size}}\\
    \subfigure[]{\includegraphics[width=0.8\columnwidth]{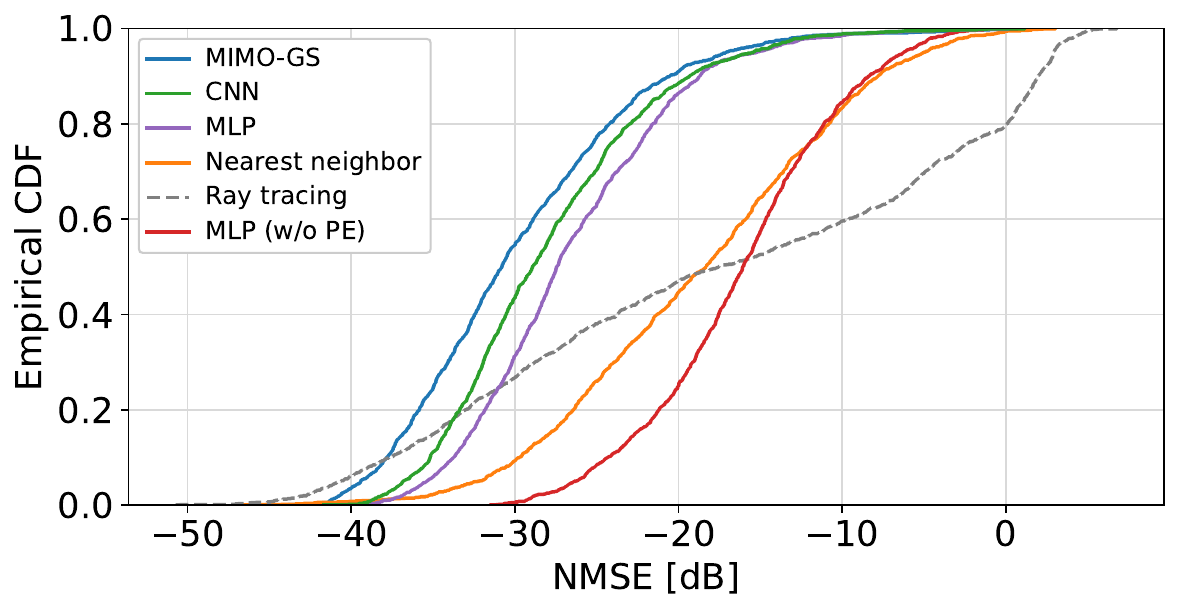}\label{Fig:nmse_cdf}}
    \caption{Rendering accuracy on the DeepMIMO dataset. (a) NMSE against the fraction of training locations with the test set fixed, where a fraction of $1$ corresponds to the $5{,}792$ training locations. (b) Empirical CDF of the per-location NMSE.}
    \label{Fig:nmse}
\end{figure}



\begin{figure}[t]
 \renewcommand{\figurename}{Fig.}
    \centering
    \includegraphics[width=0.85\columnwidth]{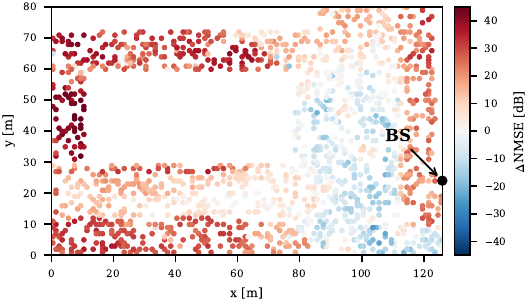}
    \caption{Per-location NMSE gap $\Delta\mathrm{NMSE}=\mathrm{NMSE}_{\mathrm{RT}}-\mathrm{NMSE}_{\mathrm{MIMO\text{-}GS}}$ over the test locations on the DeepMIMO dataset.}
    \label{Fig:error_map}
\end{figure}

The rendered beam pair power maps at two representative locations exhibit distinct spatial profiles, as shown in Fig.~\ref{Fig:samples_deepmimo}. At the first location the power spreads across several receive beams, indicating a path whose arrival is not confined to one receive direction, whereas at the second location it concentrates on a few beam pairs with a weaker secondary component. Both maps carry a regular pattern of weak vertical stripes, the sidelobes of the transmit array response at the departure angles of the paths, and the rendering reproduces these together with the positions and relative strengths of the dominant clusters.

Fig.~\ref{Fig:nmse_size} shows the NMSE as the training set is randomly subsampled with the test set fixed. MIMO-GS is the most accurate at every fraction, ahead of the CNN by $3.2$ to $5.2$~dB with the largest gap at the fewest observations. The MLP with Fourier features follows the CNN closely, whereas without them it stays near $-12$~dB, since raw coordinates bias a network toward slowly varying functions. Even with them, the MLP and the CNN fall behind, as they map each position to its map with no structure relating the maps at different positions. MIMO-GS instead shares its primitives across all locations and derives each map through the projection geometry, keeping its rendering accurate with sparser observations.


The empirical cumulative distribution function (CDF) of the per-location NMSE is shown in Fig.~\ref{Fig:nmse_cdf}. The MIMO-GS curve lies to the left over most of the range, with a median of $-30.8$~dB against $-28.9$ and $-27.4$~dB for the CNN and the MLP, $-18.3$~dB for the nearest neighbor, and $-17.5$~dB for ray tracing. The NMSE of the CNN and the MLP exceeds $-10$~dB at only $1.2\%$ and $1.5\%$ of the locations, as for MIMO-GS at $1.2\%$, against $16.7\%$ for the nearest neighbor and $40.4\%$ for ray tracing. Their gap to MIMO-GS is therefore a uniform shift rather than failures at particular locations, since they must reproduce the angular structure of each map from the position alone, whereas MIMO-GS places the dominant clusters by projecting primitives shared across all locations onto the beam grid. Ray tracing has the widest spread. Its NMSE falls below $-37.7$~dB at the best $10\%$ of the locations, where the map follows the direct geometry, and exceeds $0$~dB at $20.3\%$, where the replica built from public map data with generic materials misplaces the dominant pairs.


Fig.~\ref{Fig:error_map} plots the gap $\Delta\mathrm{NMSE}=\mathrm{NMSE}_{\mathrm{RT}}-\mathrm{NMSE}_{\mathrm{MIMO\text{-}GS}}$ between the errors of the two methods against the ground truth maps. The gap is positive wherever MIMO-GS is more accurate than ray tracing, which holds at $74.7\%$ of the locations. The two methods are within a few decibels of each other inside $60$~m of the BS, where the beam pair power map is dominated by the direct geometry between the two arrays that a replica specifies correctly. Beyond $60$~m the gap grows to $23$~dB and MIMO-GS is the more accurate at over $98\%$ of the locations, since the map there draws increasingly on reflected and diffracted contributions, each depending on the orientation and the material of the surface it strikes, so an approximate replica shifts the predicted angles and moves the dominant beam pair to a different entry.

\begin{table}[t]
\centering
\caption{Comparison of prerequisites and rendering time}
\label{table_cost}
\setlength{\tabcolsep}{3pt}
\begin{tabular}{l l c c}
\hline
Method & Prerequisite & \# of params & Time/location \\
\hline
Ray tracing & Calibrated 3D replica & --- & 2.41 ms \\
MIMO-GS & Channel observations & 557k & 0.44 ms \\
\hline
\end{tabular}
\end{table}

As summarized in Table~\ref{table_cost}, MIMO-GS renders a map in $0.44$~ms per location from a scene of $25{,}000$ primitives, i.e., $50{,}000$ Gaussians, with $557$k parameters, against $2.41$~ms for ray tracing, about $5.5$ times faster. Ray tracing computes propagation paths through the environment, so its cost scales with the site complexity and the number of interactions, whereas MIMO-GS requires only one forward pass of the gain network per primitive, followed by closed-form projections evaluated in parallel on the beam grid, and needs no calibrated 3D replica.

\subsection{Rendering on Measured Channels}

\begin{figure}[t]
 \renewcommand{\figurename}{Fig.}
    \centering
    \includegraphics[width=0.8\columnwidth]{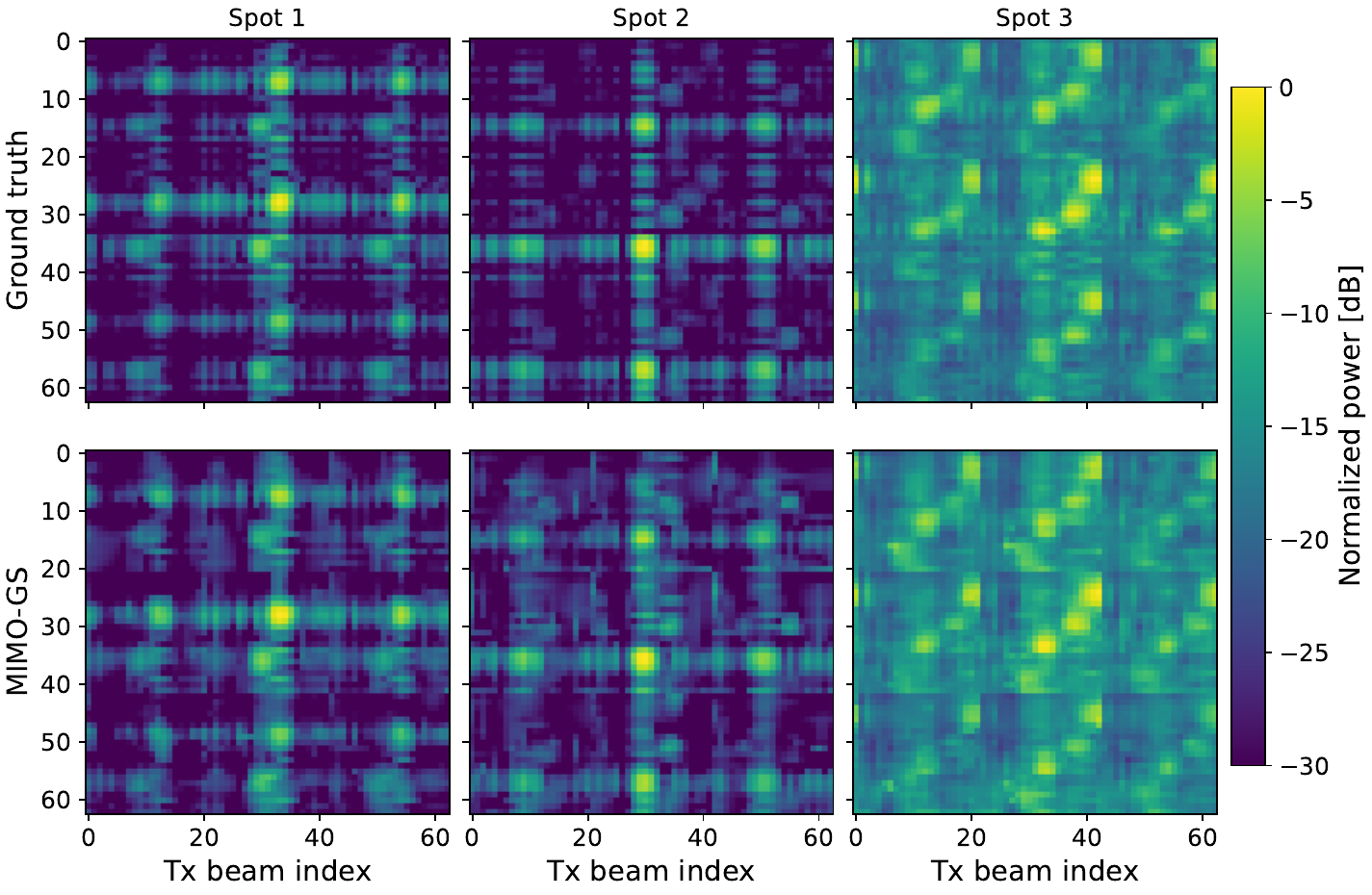}
    \caption{Ground truth and rendered beam pair power maps at three test locations on the measured indoor dataset.}
    \label{Fig:measured_samples}
\end{figure}


We next evaluate MIMO-GS on channels measured in an indoor office at $60.48$~GHz with a sounding bandwidth of $600$~MHz and $2\times8$ phased arrays at both ends~\cite{scazzoli2026dataset}. The transmitter is fixed in an adjacent corridor against the wall separating it from the room and oriented toward the doorway, as shown in Fig.~\ref{fig:scenarios}(b). The receiver measures at $350$ locations on a $0.15$~m grid inside the room, partitioned into disjoint training and test sets as for the DeepMIMO scene. The beam domain of this dataset is the steering codebook of the sounder rather than the DFT basis of the arrays. Both sides sweep the same $63$ beams over $21$ azimuth and $3$ elevation directions, so the number of beams is decoupled from the number of antennas and the beams are not mutually orthogonal. Rendering on this domain requires only relocating the beam centers $\{\mathbf{b}^{\mathrm{r}}_m\}$ and $\{\mathbf{b}^{\mathrm{t}}_n\}$ in \eqref{5_rx_raw_weight} and \eqref{5_tx_raw_weight} from the DFT grid to the codebook steering directions, and the splatting rule in \eqref{5_beam_pair_rendering} carries over unchanged. The observed map at each location is obtained from a single beam sweep, with each pair power accumulated from the channel impulse response over the delay window. This accumulation equals the average of the pair power across the sounding band and thus estimates the expectation in \eqref{2_power_map}, and measurement noise together with unresolved diffuse components remains as a residual specific to each location. As shown in Fig.~\ref{Fig:measured_samples}, at the first two locations, a few strong paths coupled through the doorway concentrate the power on narrow clusters of beam pairs, whereas the third lies in a deep shadow with no dominant path and components of comparable strength spread over the beam pair grid. The rendering reproduces both profiles. Narrow projected covariances place the concentrated clusters, and the diffuse profile is captured through the superposition of primitives with broad projected supports.

\begin{table}[t]
\centering
\caption{Rendering accuracy on the measured indoor dataset}
\label{table_measured}
\setlength{\tabcolsep}{4pt}
\begin{tabular}{l c c c c c}
\hline
Method & NMSE [dB] & Top-1 & Top-4 & Top-8 & $C_4$ \\
\hline
CNN & $-6.06$ & 0.37 & 0.69 & 0.63 & 0.79 \\
MLP & $-6.37$ & 0.55 & 0.68 & 0.65 & 0.80 \\
Nearest neighbor & $-8.06$ & 0.43 & 0.60 & 0.65 & 0.76 \\
MIMO-GS & $\mathbf{-12.36}$ & $\mathbf{0.64}$ & $\mathbf{0.81}$ & $\mathbf{0.80}$ & $\mathbf{0.92}$ \\
\hline
\end{tabular}
\end{table}

Table~\ref{table_measured} reports the rendering accuracy on the measured dataset. MIMO-GS attains an NMSE of $-12.36$ dB and the highest Top-$K$ and $C_4$ scores. The measured maps vary over far shorter distances, since the doorway and the reflectors that carry the propagation lie only a few meters from the receiver, so the dominant beam pairs shift within one or two grid steps and the nearest neighbor degrades. The MLP and the CNN fall further behind, at $-6.37$ and $-6.06$~dB. With only a few hundred training locations, they memorize the training maps, as they map each position to its map with no structure relating the maps at different positions. MIMO-GS instead fits one primitive set to all training locations, and the projection geometry carries this rapid variation by shifting the beam associations of each primitive with the location. Top-$1$ is the most sensitive to the diffuse profiles, where several pairs have nearly equal power and a small error changes which pair is the strongest, while Top-$4$, Top-$8$, and $C_4$ show that the dominant beam subspace is captured.

\subsection{Beam Management with the Rendered Maps}

\begin{figure}[t]
 \renewcommand{\figurename}{Fig.}
    \centering
    \includegraphics[width=0.85\columnwidth]{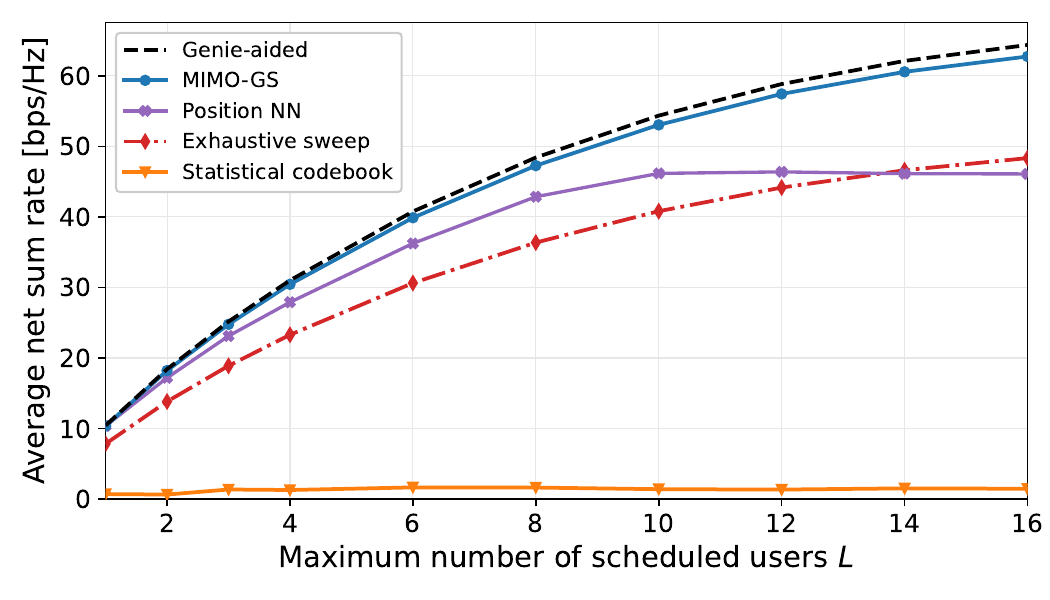}
    \caption{Net sum rate versus the maximum number of scheduled users $L$ ($U=64$ candidates, SNR $=20$~dB, $T_\mathrm{B}=1024$, $\tau_\mathrm{RS}=1$).}
    \label{Fig:net_rate_user}
\end{figure}

\begin{figure}[t]
 \renewcommand{\figurename}{Fig.}
    \centering
    \includegraphics[width=0.85\columnwidth]{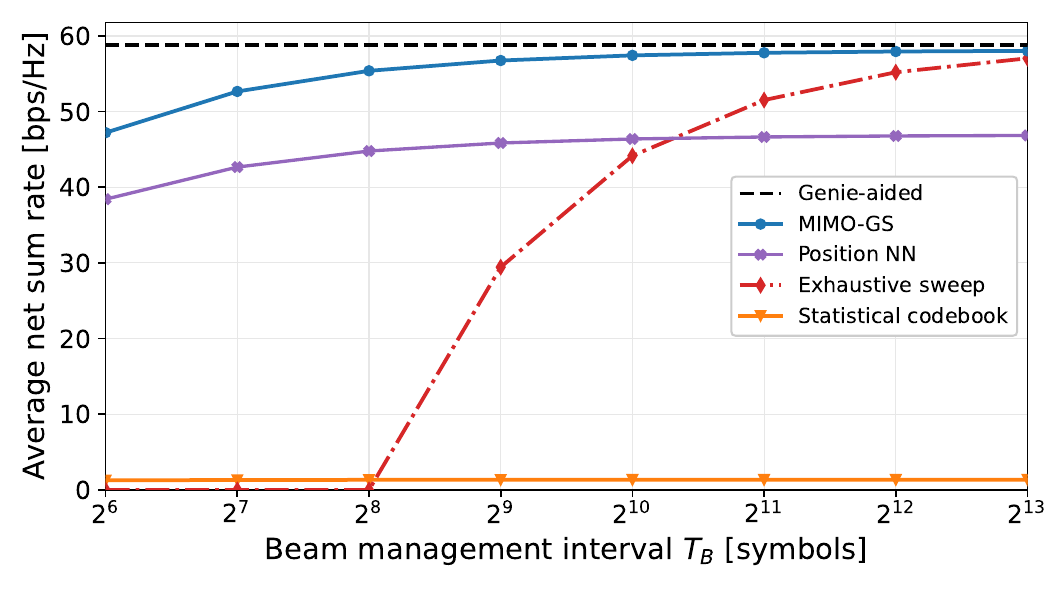}
    \caption{Net sum rate versus the beam management interval $T_\mathrm{B}$ ($U=64$, $L=12$, SNR $=20$~dB, $\tau_\mathrm{RS}=1$).}
    \label{Fig:net_rate_time}
\end{figure}

The beam management performance is evaluated on the DeepMIMO scene with the net sum rate in \eqref{6_mu_rate}. Since the maps carry only beam pair powers, the expectation in \eqref{6_mu_rate} is evaluated over complex beamspace channel realizations at each test location. The SNR is defined as $P/\sigma_z^2$ with the maps and channels rescaled to unit peak power at the location of median peak power. MIMO-GS runs Algorithm~\ref{alg_mu_sched} on its rendered maps, while the genie-aided benchmark runs it on the ground truth maps with the overhead factor in~\eqref{6_mu_rate} removed. Since the renderer is trained on peak-normalized targets and produces the distribution over the beam pairs, the peak power of each candidate is set to the mean over its five nearest training locations. The reported rates are averaged over $500$ candidate sets of $U=64$ UEs drawn uniformly from the test locations, with the CSI-RS overhead fixed at $\tau_\mathrm{RS}=1$ symbol. The rendered maps are compared with the following baselines.

\begin{itemize}
    \item Statistical codebook: users drawn at random and beams assigned from the average map over the training set
    \item Exhaustive sweep: CSI-RS on all $N_\mathrm{t}$ transmit beams, from which the long-term map of each user is measured, with users and beams then allocated by Algorithm~\ref{alg_mu_sched}
    \item Position NN: a fully connected neural network (NN) that maps the UE position to a transmit beam \cite{morais2023position}, trained on the same observation set
\end{itemize}

The first two baselines produce a map and therefore run Algorithm~\ref{alg_mu_sched} unchanged. Position NN returns a beam index rather than a map, so the algorithm does not apply. A user is scheduled on its predicted beam when that beam is free and no beam adjacent to it is in use.

Fig.~\ref{Fig:net_rate_user} shows the net sum rate against the maximum number of scheduled users. MIMO-GS attains 97 to 99\% of the genie-aided benchmark over the whole range and reaches $57.4$~bps/Hz at $L = 12$. Position NN coincides with it at $L=1$, where a single stream carries no interference, and falls behind as the users accumulate, since it does not control the interference as effectively. It saturates near $46$~bps/Hz beyond $L=10$, so the gap to MIMO-GS widens from $5.6\%$ at $L=2$ to $19.2\%$ at $L=12$, and the exhaustive sweep overtakes Position NN at $L=14$ despite its overhead. For the exhaustive sweep, the prefactor $1-|\mathcal{S}|\tau_\mathrm{RS}/T_\mathrm{B}$ in \eqref{6_mu_rate} is replaced by $1-N_\mathrm{t}\tau_\mathrm{RS}/T_\mathrm{B}$, which costs a quarter of the rate at $T_\mathrm{B} = 1024$, so the sweep stays below the genie-aided benchmark by that factor while its spectral efficiency matches the benchmark's.

Fig.~\ref{Fig:net_rate_time} shows the net sum rate against the beam management interval. The exhaustive sweep measures all $N_\mathrm{t}$ beams before it can schedule, so its prefactor is nonpositive and its net rate is zero for $T_\mathrm{B}\le N_\mathrm{t}\tau_\mathrm{RS}=256$, where the interval ends before the codebook has been covered. MIMO-GS sounds the $|\mathcal{S}|$ scheduled beams regardless of $T_\mathrm{B}$ and retains $47.2$~bps/Hz at $T_\mathrm{B}=64$, where the exhaustive sweep returns nothing, and $97.6\%$ of the genie-aided benchmark at $T_\mathrm{B}=1024$. Position NN pays the same overhead but stays $19\%$ below MIMO-GS over the whole range, since its spectral efficiency is degraded by the interference.

\section{Conclusion}

This paper formulated spatial MIMO channel rendering as learning a latent channel scene from channel observations, with the joint transmit and receive beam pair power map as the rendered quantity. MIMO-GS represents the scene with primitives that pair a receive Gaussian and a transmit Gaussian, and renders the map through dual-sided projection and beam pair splatting. The rendered maps reproduced the dominant beam pair structure at unobserved locations on a ray-traced urban scenario and over-the-air measured indoor channels. With users and transmit beams scheduled from the rendered maps and CSI-RS configured only on the scheduled beams, the net sum rate stayed within a few percent of the genie-aided benchmark at long beam management intervals and retained most of it at the shortest interval, where a full sweep yields no rate. Once learned for a given environment, the scene renders the map at any location within the scene as a lightweight inference and is reused across channel realizations.

\bibliographystyle{IEEEtran}
\bibliography{ref_mimogs}

\end{document}